\documentclass[num-refs]{wiley-article}
\usepackage{siunitx}
\usepackage{adjustbox}
\usepackage{xcolor}

\usepackage{amsmath,amssymb,amsfonts}
\usepackage{algorithmic}
\usepackage{graphicx}
\usepackage{textcomp}
\usepackage{amsmath}
\usepackage{caption}
\usepackage{multirow}
\usepackage{booktabs}

\usepackage{amsmath}
\usepackage{amssymb}
\usepackage{bm}
\usepackage{mathtools}
\usepackage{dsfont}
\usepackage{multirow}
\usepackage{leftidx}
\usepackage{textgreek}
\usepackage{newfloat}
\DeclareFloatingEnvironment[
fileext=losv,
listname=List of Supporting Information Videos,
name=Supporting Information Video S\hspace{-0.8ex},
placement=tbhp,
within=none,
]{svideo}

\usepackage{newfloat}
\DeclareFloatingEnvironment[
fileext=losf,
listname=List of Supporting Information Figures,
name=Supporting Information Figure S\hspace{-0.8ex},
placement=tbhp,
within=none,
]{sfigure}

\usepackage{newfloat}
\DeclareFloatingEnvironment[
fileext=lost,
listname=List of Supporting Information Tables,
name=Supporting Information Table S\hspace{-0.8ex},
placement=tbhp,
within=none,
]{stable}

\papertype{Full Paper}
\paperfield{Magnetic Resonance in Medicine}

\title{{Complementary Time-Frequency Domain Networks for Dynamic Parallel MR Image Reconstruction}}

\author[1,2]{Chen~Qin}
\author[3]{Jinming~Duan}
\author[2,4]{Kerstin~Hammernik}
\author[2,5]{Jo~Schlemper}
\author[6,7]{Thomas Küstner}
\author[6]{René Botnar}
\author[6]{Claudia Prieto}
\author[6]{Anthony N. Price}
\author[6]{Joseph V. Hajnal}
\author[2,4]{Daniel~Rueckert}

\affil[1]{Institute for Digital Communications, School of Engineering, University of Edinburgh, Edinburgh, United Kingdom}
\affil[2]{Department of Computing, Imperial College London, London, United Kingdom}
\affil[3]{School of Computer Science, University of Birmingham, Birmingham, United Kingdom}
\affil[4]{Institute for AI and Informatics, Klinikum rechts der Isar, Technical University of Munich, Munich, Germany}
\affil[5]{Hyperfine Research Inc., Guilford, CT, USA}
\affil[6]{School of Biomedical Engineering and Imaging Sciences, King's College London, St. Thomas' Hospital, London, United Kingdom}
\affil[7]{Department of Diagnostic and Interventional Radiology, Medical Image and Data Analysis, University Hospital of Tuebingen, Tuebingen, Germany}

\corraddress{Chen Qin, PhD, 
Institute for Digital Communications, School of Engineering, University of Edinburgh, Edinburgh, EH9 3JL, United Kingdom
}
\corremail{Chen.Qin@ed.ac.uk}

\fundinginfo{EPSRC Programme Grant: EP/P001009/1}

\runningauthor{Qin et al.}
\newif\ifisresponse
\isresponsefalse % with reviewer comments and annotations

\usepackage{marginnote}
\usepackage{color}
\definecolor{changedcolorEditor}{rgb}{0.2,0,0}
\definecolor{changedcolorRone}{rgb}{0,0.26,0.15}
\definecolor{changedcolorRtwo}{rgb}{0.48,0.07,0.07}
\definecolor{changedcolorRthree}{rgb}{0,0,0.5}
\definecolor{changedcolorRgen}{rgb}{0.4,0.4,0.4}

\usepackage[draft]{changes}
\ifisresponse

 		    \newcommand{\mdeleted}[2][None]{\deleted[]{#2}}       
\else

			\newcommand{\mdeleted}[2][None]{\deleted[]{}}	        
			\usepackage[draft]{changes}
\fi

\usepackage[capitalise]{cleveref}

\begin{document}

%TC:ignore
\maketitle

\begin{abstract}
\textbf{Purpose} 
{To introduce a novel deep learning based approach for fast and high-quality dynamic multi-coil MR reconstruction by learning a complementary time-frequency domain network that exploits spatio-temporal correlations simultaneously from complementary domains.}

\noindent\textbf{Theory and Methods} 
{Dynamic parallel MR image reconstruction is formulated as a multi-variable minimisation problem, where the data is regularised in combined temporal Fourier and spatial ($x$-$f$) domain as well as in spatio-temporal image ($x$-$t$) domain. An iterative algorithm based on variable splitting technique is derived, which alternates among signal de-aliasing steps in $x$-$f$ and $x$-$t$ spaces, a closed-form point-wise data consistency step and a weighted coupling step. The iterative model is embedded into a deep recurrent neural network which learns to recover the image via exploiting spatio-temporal redundancies in complementary domains.}

\noindent\textbf{Results} 
{Experiments were performed on two datasets of highly undersampled multi-coil short-axis cardiac cine MRI scans. Results demonstrate that our proposed method outperforms the current state-of-the-art approaches both quantitatively and qualitatively. The proposed model can also generalise well to data acquired from a different scanner and data with pathologies that were not seen in the training set.}

\noindent\textbf{Conclusion}
{The work shows the benefit of reconstructing dynamic parallel MRI in complementary time-frequency domains with deep neural networks. The method can effectively and robustly reconstruct high-quality images from highly undersampled dynamic multi-coil data ($16 \times$ and $24 \times$ yielding 15s and 10s scan times respectively) with fast reconstruction speed (2.8s). This could potentially facilitate achieving fast single-breath-hold clinical 2D cardiac cine imaging.}

% Please include a maximum of seven keywords
\end{abstract}
%TC:endignore

\section{Introduction}
Magnetic Resonance Imaging (MRI) is a widely used diagnostic modality which generates images with high spatial and temporal resolution as well as excellent soft tissue contrast. Dynamic MRI is often used to monitor dynamic processes of anatomy such as cardiac motion by acquiring a series of images at a high frame rate. However, the physics of the image acquisition process as well as physiological constraints limit the speed of MRI acquisition, and long scan time also makes it difficult to acquire images of moving structures. Thus, acceleration of the MRI acquisition is crucial to enable these clinical applications.

% Parallel imaging is frequently used to accelerate MRI data acquisition from the hardware side. In parallel imaging, data is acquired using an array of multiple independent receiver coils, which allows for a reduction in the number of phase-encoding steps during the acquisition process and enables the combination of undersampled data from each receiver coil given the additional coil information. 
{Parallel imaging (PI) techniques \cite{pruessmann1999sense,griswold2002generalized,sodickson1997simultaneous} have been widely used to accelerate MR imaging. 
% It is well known that substantial correlations in $k$-space and time exist in dynamic MRI. To accelerate the acquisition, most strategies for parallel dynamic MRI have been designed to acquire part of the desired $k$-$t$ measurements at each coil and then reconstruct the images by exploiting spatio-temporal redundancies within the data. 
% Over the years, a number of approaches have been proposed for the reconstruction of accelerated dynamic MR images. These strategies are able to speed up data acquisition without degrading image quality substantially due to the high degree of spatio-temporal correlations existing in the image series. 
%In general, these methods can be mainly divided into three categories, based on exploiting correlations in $k$-space, in time, and in both $k$-space and time \cite{tsao2003k}. 
% Parallel imaging methods are one of the groups of methods that can accelerate dynamic MRI \cite{pruessmann1999sense,griswold2002generalized,sodickson1997simultaneous,}. 
They speed up the scan time by sampling only a limited number of phase-encoding steps, and then exploiting the correlations to restore the missing information in the reconstruction phase. 
% For instance, GRAPPA (GeneRalized Autocalibrating Partially Parallel Acquisitions) \cite{griswold2002generalized} estimates the missing data directly in $k$-space, and SENSE (SENSitivity Encoding) \cite{pruessmann1999sense} recovers the uncorrupted image from a sequence of aliased images via solving a linear sensitivity equation.
% To accelerate dynamic MRI, some PI-based strategies \cite{kellman2001adaptive,breuer2005dynamic,huang2005k} have also been proposed to exploit spatio-temporal correlations in k-space and time in combination with coil sensitivity information. 
%
Compressed sensing (CS) techniques combined with PI have shown great potential in improving the image reconstruction quality and acquisition speed \cite{jung2007improved,jung2009k,otazo2010combination,lingala2011accelerated,lustig2006kt}. CS-based methods exploit signal sparsity in some specific transform domain, and recover the original image from undersampled k-space data using nonlinear reconstructions. One effective mean to exploit spatio-temporal redundancies for signal recovery in dynamic MRI is to enforce the sparsity in combined spatial and temporal Fourier ($x$-$f$) domain against consistency with the acquired undersampled k-space data, and this can be represented by methods such as  $k$-$t$ FOCUSS \cite{jung2007improved,jung2009k} and $k$-$t$ SPARSE-SENSE \cite{otazo2010combination}. The combinations of CS with low-rank in matrix completion schemes and spatio-temporal partial separability \cite{lingala2011accelerated,otazo2015low,zhao2012image} have also been proposed to exploit correlations between the temporal profiles of the voxels, e.g. $k$-$t$ SLR \cite{lingala2011accelerated}. Some more recent approaches \cite{yoon2014motion,mohsin2017accelerated} also utilised patch-based regularisation frameworks to exploit geometric similarities in the spatio-temporal domain.} However, these CS-based approaches often require careful selection of problem-specific regularisation schemes and the tuning of hyper-parameters is often non-trivial. Furthermore, the reconstruction speed of these methods is often slow due to the iterative nature of the optimisation used, and in the context of dynamic imaging, the additional time domain further increases the computational demand. 

In contrast, deep learning (DL) based reconstruction approaches have become extremely popular in recent years and have enabled progress beyond the limitations of traditional CS techniques \cite{knoll2019deep,hammernik2020machine,eo2018kiki,ye2018deep,tezcan2018mr,yang2017dagan}. {In DL methods, prior information and regularisation can be implicitly learnt from the acquired data without having to manually specify them beforehand. Additionally, image quality and reconstruction speed are improved substantially. These advances include applications in both PI \cite{hammernik2018learning,aggarwal2018modl,kwon2017parallel,cheng2018deepspirit,lonning2018recurrent,schlemper2019data,hammernik2019sigma,fuin2020multi} and dynamic MRI \cite{schlemper2018deep,schlemper2018bayesian,qin2019convolutional,hauptmann2019real,seegoolam2019exploiting,biswas2019dynamic,hammernik2019dynamic}. Most current approaches in DL for accelerated PI are based on exploiting information in a single image either in image domain \cite{hammernik2018learning,aggarwal2018modl,schlemper2019data} or in k-space domain \cite{akccakaya2019scan,han2018k,zhang2018multi}, where each image (or frame) is reconstructed independently. Examples of these include the variational network (VN) \cite{hammernik2018learning} and robust artificial-neural-networks for k-space interpolation (RAKI) etc. In accelerated dynamic MRI, one of the key ingredients is to exploit the temporal redundancies. To this end, 3D convolutional networks (Cascade CNN) \cite{schlemper2018deep} and bidirectional convolutional recurrent neural networks (CRNN) \cite{qin2019convolutional} have been proposed to exploit the temporal dependencies of dynamic sequences in spatio-temporal image domain.
% Schlemper et al. \cite{schlemper2018deep} proposed a cascaded 3D convolutional network with a data sharing (DS) layer to utilise similar information contained in neighboring $k$-space samples, and Qin et al. \cite{qin2019convolutional} designed a bidirectional convolutional recurrent neural network (CRNN) model to exploit the temporal dependencies of dynamic sequences in spatio-temporal image domain. 
Most of these DL-based approaches so far have focused on either 2D static PI or single-coil dynamic MRI, whereas only a few methods exist for dynamic parallel MRI reconstruction \cite{biswas2019dynamic,hammernik2019dynamic,kustner2020cinenet}. Thus more efficient and effective DL models for dynamic parallel MRI are highly desirable.}

{In this work, inspired by CS-based $k$-$t$ methods, we formulate the dynamic parallel MR image reconstruction as a multi-variable minimisation problem considering regularisation in both spatio-temporal and temporal frequency domains. We propose a novel end-to-end trainable deep recurrent neural network to model the iterative process resulting from the multi-variable minimisation. Specifically, the proposed DL approach alternates among four steps: (1) a signal de-aliasing step in combined spatial and temporal frequency domain ($x$-$f$) via an $xf$-CRNN; (2) a complementary de-aliasing step in spatio-temporal image domain ($x$-$t$) with an $xt$-CRNN; (3) a closed-form point-wise data consistency (DC) step and (4) a closed-form weighted coupling step which are embedded as layers in the deep neural network (DNN). Each of these steps correspond to the iterative algorithm derived from a variable splitting technique (Section \ref{Sec:vs-next}). As the proposed model exploits spatio-temporal redundancies from Complementary Time-Frequency domains for the effective image reconstruction, we term our model as CTFNet.

The main contributions of our work can be summarised as follows: Firstly, we propose a new regularisation method built on recurrent neural networks for data regularisation in complementary spatio-temporal and temporal frequency domains to fully exploit data redundancies. Though previous studies \cite{eo2018kiki,sriram2020grappanet,wang2019dimension} have shown that MR reconstruction can be performed in both k-space and image domains, it is unclear how cross-domain knowledge can be effectively utilised by DNNs in the dynamic setting, with an extra temporal dimension. To the best of our knowledge, this is the first work that investigates how complementary domain knowledge can be exploited in learning-based dynamic reconstruction. Secondly, we propose a closed-form DC layer that does not require a complex matrix inversion, and operates together with a weighted coupling layer for multi-coil images. Compared to other works \cite{hammernik2018learning,aggarwal2018modl,mardani2018deep}, our approach offers an exact update for DC, avoiding the expensive need of solving a linear system via gradient updates. This enables our approach to be computationally more efficient and simpler for implementation. Finally, we demonstrate that our approach is able to further push the undersampling rates with improved image quality against state-of-the-art CS and DL methods, as well as with a good generalisation ability to unseen data. This indicates a great potential in achieving fast single-breath-hold 2D cardiac cine imaging.}

{This work extends our preliminary conference work on single-coil dynamic MRI reconstruction \cite{qin2019k} and 2D static parallel MRI reconstruction \cite{duan2019vs}, where we explored dynamic MRI and static PI separately. In comparison to our previous work, this work presents a novel and unified end-to-end DL solution with a new formulation for dynamic parallel MRI reconstruction, which addresses a more common scenario in the use-case for clinical practice. It proposes a new way of exploiting complementary time-frequency domain information in DL. Significantly more thorough quantitative and qualitative evaluations of the proposed method including comparison, generalisation and ablation studies have been performed on multi-coil cardiac MR data with retrospective undersampling.}

\section{Theory}
\label{Sec:vs-next}
\label{Sec:problem-formulation}
\subsection{Dynamic parallel MRI model}
Assume that $\mathbf{m} \in {\mathbb{C}^N}$ is a complex-valued MR image sequence in $x$-$y$-$t$ space represented as a vector, and let $\mathbf{v}_i \in {\mathbb{C}^M}$ ($M \ll N$) denote the undersampled k-space data (in $k_x$-$k_y$-$t$ space) measured from the $i$th MR receiver coil. The data acquired from each coil thus can be represented as  
\begin{equation}
\mathbf{v}_i={\mathbf{D}} {\mathbf{F}}_\text{s} \mathbf{S}_i \mathbf{m},
\end{equation}
where ${\mathbf{F} }_{\text{s}}$ is the spatial Fourier transform matrix, ${\mathbf{D}}$ is the sampling matrix on a Cartesian grid that zeros out entries that are not acquired, and ${\mathbf{S}_i }$ is the $i$th coil sensitivity map. The reconstruction of $\mathbf{m}$ from $\mathbf{v}_i$ is an ill-posed inverse problem, where $i \in \left\{ {1,2,...,{n_c}} \right\}$ and $n_c$ denotes the number of receiver coils. {Similar to CS formulations \cite{otazo2010combination,lustig2007sparse,block2007undersampled} based on the SENSE model,  we formulate dynamic parallel MRI reconstruction as the following optimisation problem:}   
\begin{equation} \label{eq:sense}
\mathop {\min }\limits_\mathbf{m}  { {{\cal{R}}_{\text{xf}}}\left( {\mathbf{F}}_\text{t} \mathbf{m} \right) + \mu {{\cal R}_{\text{xt}}}\left( \mathbf{m} \right)+\frac{\lambda }{2}\sum\limits_{i = 1}^{{n_c}} {\| {{\mathbf{D}} {{\mathbf{F}}_\text{s}}{\mathbf{S}_i}\mathbf{m} - {\mathbf{v}_i}} \|_2^2}  } .
\end{equation}
{Here, ${\cal R}_\text{xt}$ is defined as a regularisation term on the spatio-temporal domain ($x$-$y$-$t$ space, also denoted as $x$-$t$) of the image sequence $\mathbf{m}$, similar to the spatio-temporal total variation in most CS-based approaches. To fully exploit the spatio-temporal correlations, we additionally add a regularisation term ${\cal R}_\text{xf}$ to regularise the data in the combined spatial and temporal frequency domain ($x$-$f$ space), in which ${\mathbf{F}}_\text{t}$ denotes the temporal Fourier transform. This leverages the characteristic that the signal can be sparsely represented in the temporal Fourier domain, because of the periodic cardiac motion exhibited in dynamic imaging. Previous works \cite{eo2018kiki,lustig2007sparse,tsaig2006extensions,qin2019k} have shown that data regularisation in different domains is beneficial due to the complementary information they represent, and thus here we propose to combine the regularisation terms from the complementary time and frequency domains with $\mu$ to balance between ${\cal R}_\text{xf}$ and ${\cal R}_\text{xt}$. The last term in Eq. \ref{eq:sense} enforces the data fidelity in PI, and here we formulate it as a coil-wise DC term, which aims to avoid the need to solve a linear problem inside subsequent sub-problem and also makes it simple to be embedded in an end-to-end DL framework (see following Optimisation). The model weight $\lambda$  balances between regularisation and data fidelity. }

\textbf{Optimisation:} To optimise Eq. \ref{eq:sense}, {we propose to employ the variable splitting technique \cite{ramani2010parallel,duan2019vs}} to decouple the data fidelity term and regularisation terms. Specifically, auxiliary splitting variables $\mathbf u  \in {\mathbb{C}^N} $, $\boldsymbol{\rho}   \in {\mathbb{C}^N} $ and $\{\boldsymbol{\sigma}_i  \in {\mathbb{C}^N}\}_{i=1}^{n_c}$ are introduced here, converting Eq. \ref{eq:sense} into the following equivalent form:
\begin{equation} \label{eq:varaible_splitting}
\begin{split}
& \mathop {\min }\limits_{\mathbf{m},\mathbf{u},\boldsymbol{\rho}, \boldsymbol{\sigma}_i}  { {{\cal R}_\text{xf}}\left( \boldsymbol{\rho} \right) + \mu {{\cal R}_\text{xt}}\left( \mathbf{u} \right)+\frac{\lambda }{2}\sum\limits_{i = 1}^{{n_c}} {\| {{\mathbf{D}} {{\mathbf{ F}}_\text{s}}\boldsymbol{\sigma}_{i} - {\mathbf{v}_i}} \|_2^2}  }  \; \\
& s.t.\;\mathbf{m}=\mathbf{u}, \;{\mathbf{F}}_\text{t}\mathbf m = \boldsymbol{\rho}, \; \mathbf{S}_i\mathbf{m}=\boldsymbol{\sigma}_i, \;\forall i \in \left\{ {1,2,...,{n_c}} \right\}.
\end{split}
\end{equation} 
In detail, the introduction of the first constraint $\mathbf{m}=\mathbf{u}$ decouples $\mathbf{m}$ in the regularisation term ${\cal R}_\text{xt}$ from that in the data fidelity term, and the second constraint ${\mathbf{F}}_\text{t}\mathbf m = \boldsymbol{\rho}$ enables the decoupling of ${\cal R}_\text{xf}$ from the other terms. The introduction of the third constraint $\mathbf{S}_i\mathbf{m}=\boldsymbol{\sigma}_i$ is also crucial as it allows decomposition of $\mathbf{S}_i\mathbf{m}$ from ${\mathbf{D}} {\mathbf{F}}_\text{s}{\mathbf{S}_i}\mathbf{m} $ in the data fidelity term, {which avoids the difficult dense matrix inversion} in subsequent calculations (see Eq. \ref{eq:pDC}). Using the penalty function method, Eq. \ref{eq:varaible_splitting} can be reformulated to minimise the following single cost function:
\begin{equation}\label{eq:splitsense}
% \scalebox{0.95}{
\mathop {\min }\limits_{\mathbf{m},\mathbf{u},\boldsymbol{\rho}, \boldsymbol{\sigma}_i}   {{\cal R}_\text{xf}}\left( \boldsymbol{\rho} \right) + \mu {{\cal R}_\text{xt}}\left( \mathbf{u} \right)+\frac{\lambda }{2}\sum\limits_{i = 1}^{{n_c}} {\| {{\mathbf{D}} {\mathbf{F}}_\text{s}\boldsymbol{\sigma}_{i} - {y_i}} \|_2^2} + \frac{\alpha }{2}\| {\mathbf{u} - \mathbf{m}} \|_2^2\ + \frac{\beta }{2}\| {\boldsymbol{\rho} - {\mathbf{F}}_\text{t}\mathbf{m}} \|_2^2\ + \frac{\gamma }{2}\sum\limits_{i = 1}^{{n_c}} {\| {{\boldsymbol{\sigma}_i} - {\mathbf{S}_i}\mathbf{m}} \|_2^2},
% }
\end{equation}
where $\alpha$, $\beta$ and $\gamma$ are penalty weights. To minimise Eq. \ref{eq:splitsense} which is a multi-variable optimisation problem, alternating minimisation over $\mathbf{m}$, $\mathbf{u}$, $\boldsymbol{\rho}$ and $\boldsymbol{\sigma}_i$ is performed, resulting in iteratively solving the following sub-problems:
\begin{subequations} \label{eq:subproblems}
\begin{align}
{\boldsymbol{\rho}^{k + 1}} &= \mathop {\arg \min }\limits_{\boldsymbol{\rho}} \frac{\beta }{2}\| {\boldsymbol{\rho} - {\mathbf{F}}_\text{t} {\mathbf{m}^k}} \|_2^2 + {\cal R}_\text{xf}\left( \boldsymbol{\rho} \right), \label{eq:proximal_xyf}\\ 
{\mathbf{u}^{k + 1}} &= \mathop {\arg \min }\limits_{\mathbf{u}} \frac{\alpha }{2}\| {\mathbf{u} - {\mathbf{m}^k}} \|_2^2 + \mu {\cal R}_\text{xt}\left( \mathbf{u} \right), \label{eq:proximal_xyt}\\
\boldsymbol{\sigma}_i^{k + 1} &= \mathop {\arg \min }\limits_{{\boldsymbol{\sigma}_i}} \frac{\lambda }{2} \sum\limits_{i = 1}^{{n_c}} {\| {{\mathbf{D}} {\mathbf{F}}_\text{s}{\boldsymbol{\sigma}_i} - {\mathbf{v}_i}} \|_2^2} + \frac{\gamma }{2}\sum\limits_{i = 1}^{{n_c}} {\| {{\boldsymbol{\sigma}_i} - {\mathbf{S}_i}{\mathbf{m}^k}} \|_2^2}, \label{eq:data_consistency}\\
\mathbf{m}^{k+1} &= \begin{aligned}[t]\mathop {\arg \min }\limits_\mathbf{m}   \frac{\alpha }{2}\| {\mathbf{u}^{k + 1}} - \mathbf{m} \|_2^2 +\frac{\beta }{2}\| {\boldsymbol{\rho}^{k+1} - {\mathbf{F}}_\text{t} \mathbf{m}} \|_2^2
+\frac{\gamma }{2}\sum\nolimits_{i = 1}^{{n_c}}{\| {\boldsymbol{\sigma}_i^{k + 1} - {\mathbf{S}_i}\mathbf{m}} \|_2^2}.\end{aligned} \label{eq:weighted_combination}
\end{align}
\end{subequations}
Here, $k \in \{0,1,2,...,n_{it}-1\}$ denotes the $k$th iteration and $\mathbf{m}^0$ is the zero-filled reconstruction as an initialisation. An optimal solution ($\mathbf{m}^*$) can be found by iterating over $\boldsymbol{\rho}^{k+1}$,${\mathbf{u}^{k + 1}}$, $\boldsymbol{\sigma}_i^{k+1}$ and ${\mathbf{m}^{k + 1}}$ until convergence or reaching the maximum number of iterations $n_{it}$. 

Specifically, Eq. \ref{eq:proximal_xyf} and Eq. \ref{eq:proximal_xyt} are the proximal operators of the combined temporal Fourier and spatial domain prior ${\cal R}_\text{xf}$ and the spatio-temporal image domain prior ${\cal R}_\text{xt}$ respectively. Eq. \ref{eq:data_consistency} is a coil-wise data consistency step in PI (pDC), which imposes the consistency between the acquired k-space measurements and the reconstructed data. A closed-form solution for Eq. \ref{eq:data_consistency} can be derived as:
\begin{equation}\label{eq:pDC}
\boldsymbol{\sigma}_i^{k + 1} = {{\mathbf{F}}_\text{s}^\text{H}}( {{{( {\lambda {{\mathbf{D}} ^\text{T}}{\mathbf{D}}  + \gamma \mathbf{I} } )}^{-1}}( {\gamma {\mathbf{F}}_\text{s}{\mathbf{S}_i}{\mathbf{m}^k} + \lambda{{\mathbf{D}} ^\text{T}}{\mathbf{v}_i}} )}),
\end{equation}
in which ${{\mathbf{F}}_\text{s}^\text{H}}$ is the conjugate transpose of $\mathbf{F}_\text{s}$ and $\mathbf{I}$ is the identity matrix. Similarly, by optimising Eq. \ref{eq:weighted_combination}, we obtain the following solution:
\begin{equation}\label{eq:WAL}
\begin{split}
{\mathbf{m}^{k + 1}} = &(\alpha\mathbf{I}+\beta\mathbf{I}+\gamma\sum\nolimits_{i = 1}^{{n_c}} {\mathbf{S}_i^\text{H}} \mathbf{S}_i)^{-1}({\alpha {\mathbf{u}^{k + 1}} + \beta {{{\mathbf{F}}_t^\text{H}}\boldsymbol{\rho}^{k + 1}} +\gamma \sum\nolimits_{i = 1}^{{n_c}} {\mathbf{S}_i^\text{H}} \boldsymbol{\sigma}_i^{k + 1}}),
\end{split}
\end{equation}
where $\mathbf{S}_i^\text{H}$ is the conjugate transpose of $\mathbf{S}_i$. This can be regarded as a {weighted coupling (wCP)} of the results obtained from Eq. \ref{eq:proximal_xyf}, Eq. \ref{eq:proximal_xyt} and Eq. \ref{eq:data_consistency}. {In particular, it can be seen that both Eq. \ref{eq:pDC} and Eq. \ref{eq:WAL} are closed-form solutions and can be computed in a point-wise manner due to the inversion of diagonal matrices. This avoids iterative gradient updates and thus enables fast reconstruction speed in comparison to conjugate gradient-based approaches \cite{aggarwal2018modl,biswas2019dynamic,ramani2010parallel}. 
% In contrast to other approaches that employ conjugate gradient algorithm to solve the data fidelity step \cite{aggarwal2018modl,biswas2019dynamic,ramani2010parallel}, our solution allows exact steps \cite{ramani2010parallel} and avoids iterative gradient updates, which enables faster reconstruction speed and also makes it appealing for implementation in embedding them as layers into end-to-end DNNs. 
}

% An initial solution to these subproblems can be are derived as follows
% \begin{subequations} \label{eq:solutions}
% \small
% \begin{align}
% {\rho^{k + 1}} &= yf\text{-CRNN}({m^k})\\
% {u^{k + 1}} &= \text{CRNN}_s({m^k})\\
% x_i^{k + 1} &= {{\cal F}_s^{H}}( {{{( {\lambda {{\cal D} ^T}{\cal D}  + \gamma I } )}^{-1}}( {\gamma {\cal F}_s{S_i}{m^k} + \lambda{{\cal D} ^T}{y_i}} )})\\
% {m^{k + 1}} &= \frac{1}{\alpha + \beta + \gamma}({\alpha {u^{k + 1}} + \beta {{{\cal {F}}_t^H}\rho^{k + 1}} +\gamma \sum\nolimits_{i = 1}^{{n_c}} {S_i^H} x_i^{k + 1}})
% \end{align}
% \end{subequations}

\subsection{CTFNet for dynamic parallel MRI reconstruction}
\begin{figure}[!t]
\centering
\includegraphics[width=0.9\linewidth]{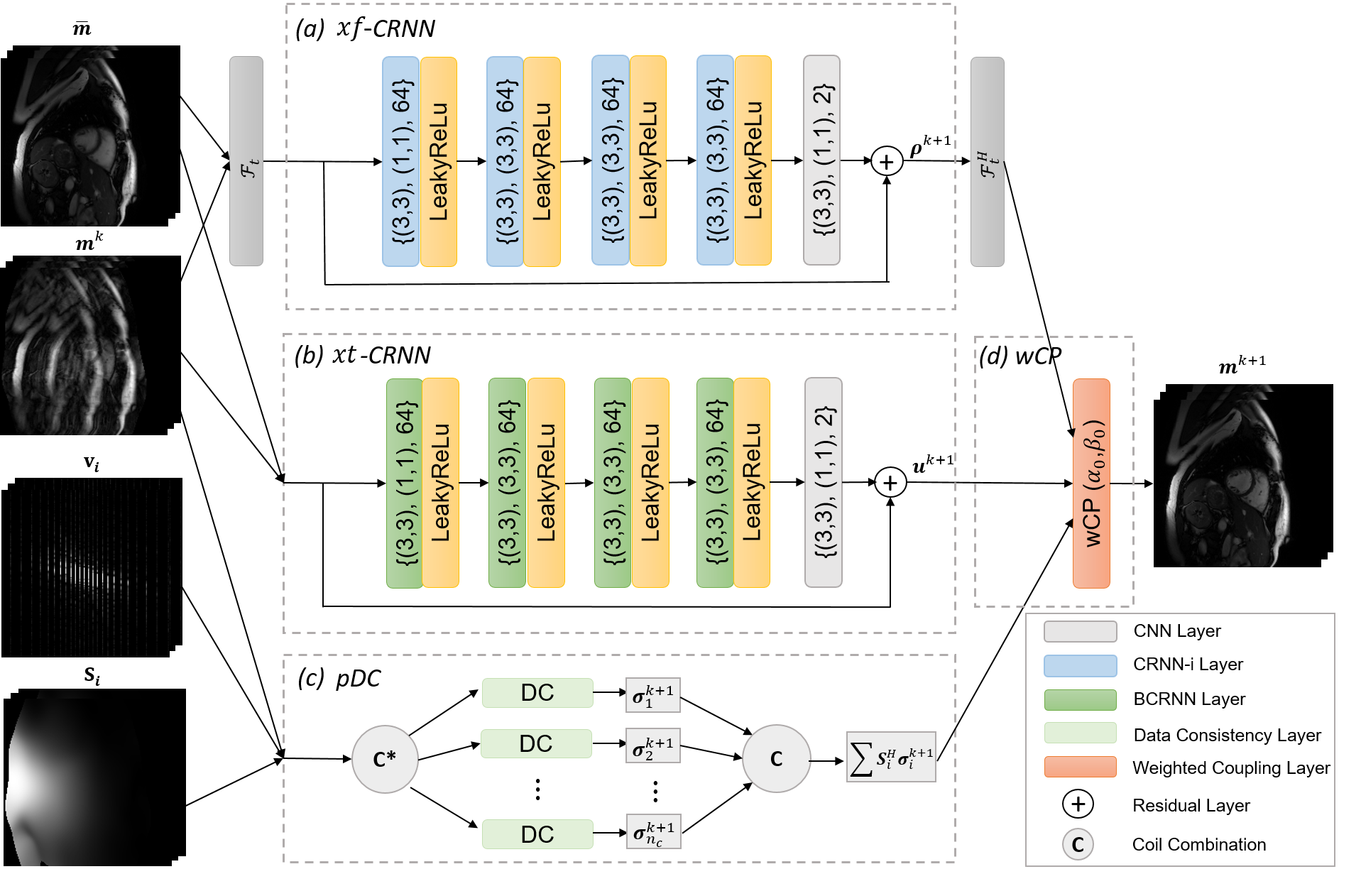}
\caption {An illustrative diagram of the proposed CTFNet at a \textbf{single iteration}.  Each component is corresponding to each subequation in Eq. \ref{eq:compact_represent} respectively. (a) Network architecture for $xf$-CRNN;
% , which is composed of 4 layers of CRNN-i and 1 layer of 2D CNN with a residual connection from the baseline estimate
 (b) network architecture for $xt$-CRNN;
% , where a variation of architecture \cite{qin2019convolutional} is employed which consists of 4 layers of BCRNN evolving over both temporal and iteration dimensions, 1 layer of 2D CNN and a residual connection;} 
(c) PI data consistency (pDC) layer; (d) weighted coupling (wCP) layer. Numbers inside CNN, CRNN-i and BCRNN layers indicate \{kernel size, dilation factor, number of filters\}. 
% {We used dilated 2D convolutions with kernel size $3 \times 3$ and dilation factor $3 \times 3$. The number of input and output channels of the network was 2, representing the real and imaginary part of the complex-valued data. 
Note that features learned at each iteration is propagated along iteration steps via the hidden-to-hidden connections in CRNN and BCRNN units. For mathematical notations, please refer to Eq. \ref{eq:compact_represent}. }
\label{fig:architecture}
\end{figure}

{Based on the model formulation in Eq. \ref{eq:subproblems}, we propose to embed the iterative reconstruction process into a DL framework to further improve the reconstruction quality with faster reconstruction speed and higher acceleration rates. Specifically, we propose a complementary time-frequency domain network (CTFNet) for the dynamic parallel MRI reconstruction to exploit the spatio-temporal correlations in complementary spatio-temporal and temporal frequency domains. Our model consists of four core components: (1) an $xf$-CRNN to implicitly learn the regularisation from the training data itself and perform the iterative de-aliasing in $x$-$f$ domain, corresponding to Eq. \ref{eq:proximal_xyf}; (2) an $xt$-CRNN similarly as the learning-based proximal operator in the spatio-temporal image domain, corresponding to Eq. \ref{eq:proximal_xyt}; (3) a pDC layer that performs coil-wise DC in PI (Eq. \ref{eq:data_consistency}); and (4) a wCP layer that is naturally derived from Eq. \ref{eq:weighted_combination} and performs the weighted coupling. An illustrative diagram of the proposed model is shown in Fig. \ref{fig:architecture}. Note that the iterative reconstruction process as stated in Eq. \ref{eq:subproblems} is modelled via the convolutional recurrent neural networks (CRNN) with recurrence over iterations. Details of each component of our network is explained hereafter.}

% Particularly, instead of explicitly imposing the regularisation terms on the data as shown in Eq. \ref{eq:proximal_xyf} and Eq. \ref{eq:proximal_xyt}, here we propose to implicitly learn the regularisation from the training data itself by leveraging deep neural networks in both domains. In particular, we employ the convolutional recurrent neural networks (CRNN) as the learning-based proximal operators ($xf$-CRNN and $im$-CRNN) to formulate the iterative reconstruction process in a recurrent way, which has been shown to be an effective technique in dynamic MR reconstruction \cite{qin2019convolutional}. Additionally, Eq. \ref{eq:data_consistency} and Eq. \ref{eq:weighted_combination} can also be correspondingly formulated as a pDC layer and a WA layer in the neural network (Eq. \ref{eq:pDC} and Eq. \ref{eq:WAL}) respectively.

\subsubsection{${xf}$-CRNN}
% Consider a Cartesian $k$-space trajectory where $k_x$ denotes the phase encoding direction and $k_y$ denotes the readout direction. 
% It is known that the temporal Fourier transform is effective in sparsifying the signal when the image follows periodic motion.
% From the perspective of compressed sensing, the dynamic MR image reconstruction problem is commonly formulated as exploiting the sparsity of the unknown spectral signal $\boldsymbol{\rho}$ in $x$-$f$ domain\cite{jung2007improved,jung2009k,otazo2010combination,lustig2006kt}, such that, 
% \begin{equation}\label{eq:focuss}
% \text{min} \ ||{\boldsymbol{\rho}}||_{1}, \quad s.t.\ ||\mathbf{v}_i-\mathcal{F} \mathbf{S}_i\boldsymbol{\rho}||_{2}\leq \epsilon,
% \end{equation}
% where $\epsilon$ denotes the noise level and $\mathcal{F}$ is the Fourier transform along the $x$-$f$ direction. This formulation is also corresponding to $\mathcal{R}_t(\boldsymbol{\rho})$ in Eq. \ref{eq:varaible_splitting}. 
{Corresponding to Eq. \ref{eq:proximal_xyf}, we first propose to exploit the spatio-temporal correlations in the combined temporal Fourier and spatial domain. Instead of explicitly imposing the regularisation term on the data such as in conventional CS-based methods, here we propose to implicitly learn the regularisation from the training data itself by leveraging DNNs in the $x$-$f$ domain. Specifically, motivated by some of the CS-based $k$-$t$ methods such as $k$-$t$ FOCUSS \cite{jung2007improved,jung2009k}, where its solution to the underdetermined inverse problem can be expressed as the form that consists of a baseline signal $\bar{\rho}$ together with its residual encoding $(\boldsymbol{\rho}^{k}-\bar{\boldsymbol{\rho}})$ for the $k+1$-th estimate of the $x$-$f$ signal $\boldsymbol{\rho}^{k+1}$,
% In k-t FOCUSS \cite{jung2007improved,jung2009k}, the underdetermined inverse problem was solved via a sparse reconstruction algorithm called  FOCal Underdetermined System Solver (FOCUSS) \cite{gorodnitsky1995neuromagnetic}, and its solution to Eq. \ref{eq:focuss} can be expressed as the form that consists of a baseline signal $\bar{\rho}$ together with its residual encoding for the $k+1$-th estimate of the $x$-$f$ signal $\boldsymbol{\rho}^{k+1}$:
% \begin{equation}\label{FOCUSS}
%     \boldsymbol{\rho}^{k+1} = \bar{\boldsymbol{\rho}} + \text{FOCUSS}(\boldsymbol{\rho}^{k}-\bar{\boldsymbol{\rho}}, \boldsymbol{\rho}^{k}),
% \end{equation}
we propose to formulate our $x$-$f$ domain reconstruction as
\begin{equation}\label{eq:xf_crnn_rho}
    \boldsymbol{\rho}^{k+1} = \bar{\boldsymbol{\rho}} + xf\text{-CRNN}(\boldsymbol{\rho}^{k}-\bar{\boldsymbol{\rho}}).
\end{equation}
% In detail, FOCUSS in Eq. \ref{FOCUSS} denotes the FOCal Underdetermined System Solver \cite{gorodnitsky1995neuromagnetic} for the sparse reconstruction algorithm, and its mathematical form \cite{jung2007improved,jung2009k} is omitted in this work for simplicity.

Particularly, in our formulation of Eq. \ref{eq:xf_crnn_rho}, different from model-based \cite{tsao2003k} or compressed sensing \cite{jung2007improved,otazo2010combination} algorithms, we employ a stack of convolutional layers to estimate the missing data based on other available points, typically within its vicinity in $x$-$f$ space. To fully exploit the spatio-temporal redundancies, we use the temporal average of a sequence as the $x$-$f$ baseline signal $\bar{\boldsymbol{\rho}}$, and thus $xf$-CRNN learns to reconstruct residuals of the temporal frequencies with respect to the temporal average (direct current) values. This makes the residual energy much sparser and enables the network to focus more on the dynamic patterns of the signals with less efforts in reconstructing static background regions.} In contrast to $k$-$t$ FOCUSS implementation where sparsity was exploited for each coil separately, the proposed approach exploits the joint information in the multi-signal ensemble that represents the combination from all coils. This has been shown to be effective in reducing the number of required samples per coil and providing increased acceleration capability \cite{otazo2010combination}. {Furthermore, different from our previous work in \cite{qin2019k}, we propose to model the iterative reconstruction process in $x$-$f$ domain with the recurrent neural network (CRNN-i \cite{qin2019convolutional}) where recurrence is evolving over iterations via hidden-to-hidden connections and the trainable network parameters are shared across sequential iteration steps.} 
% This enables effective information propagation along iterations to achieve good reconstruction with fewer number of parameters than independently parameterised networks \cite{qin2019convolutional}.

\begin{figure}[!t]
\centering
{\includegraphics[width=0.7\linewidth]{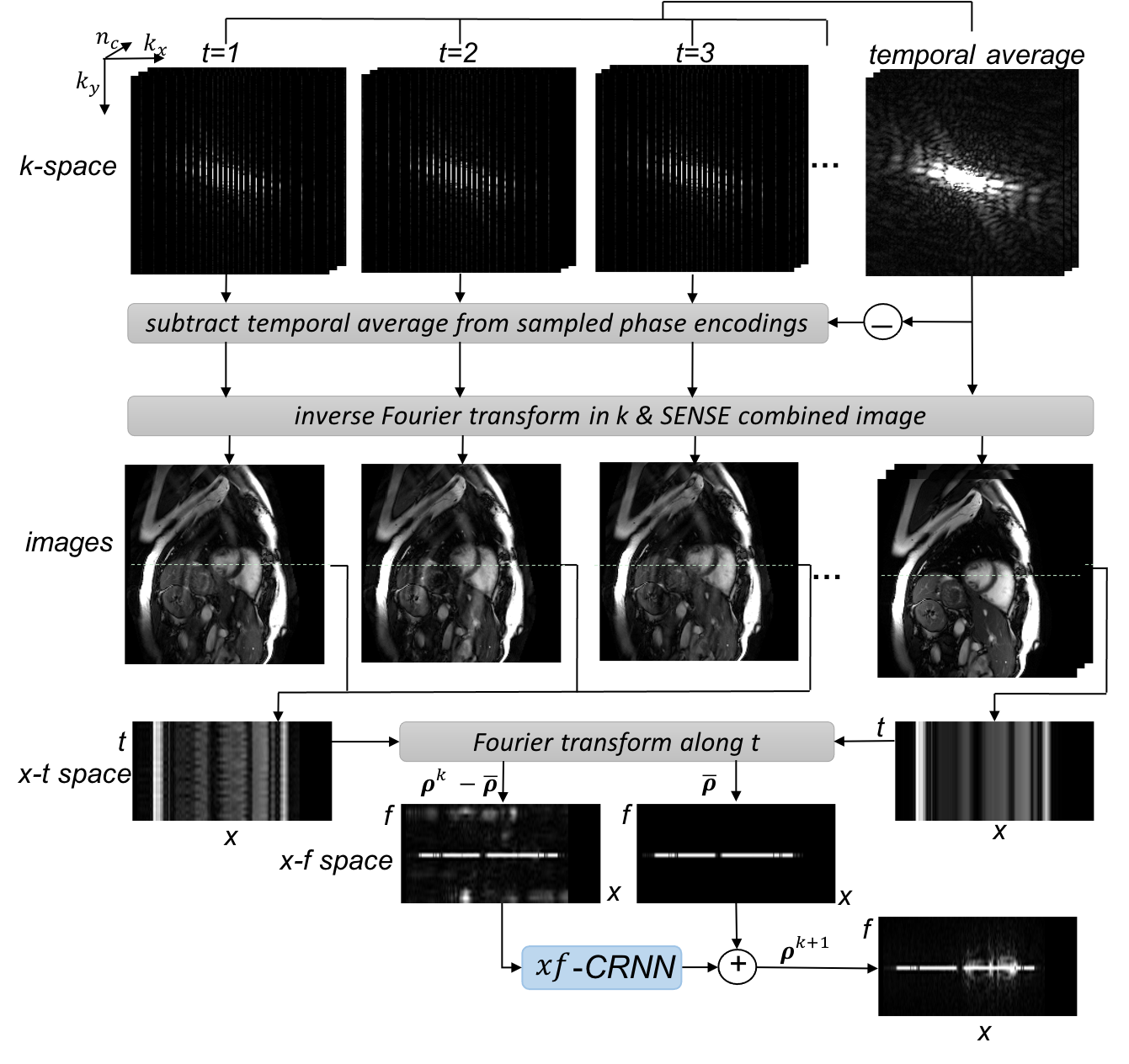}}
\caption{The $x$-$f$ transform and reconstruction diagram for a single iteration in the combined spatial and temporal frequency space. {In detail, the $x$-$f$ transform layer receives input from multi-coil $k$-$t$ space data. The acquired multi-coil k-space data is firstly averaged along $t$ to yield a temporal average for each coil separately. At iteration $k$, the temporally averaged data is subtracted from corresponding coil data at each time frame,
and the subtracted data and temporally averaged data from multi-coils are then inverse Fourier transformed and sensitivity-combined back to image space. This  yields a sequence of aliased images and a temporally averaged sequence (Eq. \ref{temporal_average}). Each frequency-encoding position of the coil-combined images is then processed separately hereafter. The image rows from aliased images or baseline images are gathered and  temporal Fourier transformed along $t$ to yield an $x$-$f$ image, corresponding to $\boldsymbol{\rho}^{k}-\bar{\boldsymbol{\rho}}$ and $\bar{\boldsymbol{\rho}}$ respectively. These signals are then fed as inputs to $xf$-CRNN for $x$-$f$ space reconstruction (Eq. \ref{eq:xf_crnn_rho} and Eq. \ref{xfcnn}).}}
\label{fig:flow}
\end{figure}

The illustrative diagram of $x$-$f$ reconstruction is shown in Fig. \ref{fig:flow}. Specifically, we formulate the $k$-$t$ to $x$-$f$ transformation process in PI as an {{${x}$-${f}$ transform layer}} in the network. The $x$-$f$ transform layer receives input from multi-coil $k$-$t$ space data, {and then transform it to $x$-$f$ space as inputs to $xf$-CRNN. Details of the process are illustrated and explained in Fig. \ref{fig:flow}.} {Note that the value range of the direct current component of the undersampled data in x-f space is lower than that of the temporal average, therefore after subtraction, the direct current component still remains but with a different value range from the temporal average. This also means that the subtracted data in image space can look similar to the temporal average but with a lower intensity range and aliasing artefacts.}
% In detail, the $x$-$f$ transform layer receives input from multi-coil $k$-$t$ space data. The acquired multi-coil $k$-space data is firstly averaged along $t$ to yield a temporal average for each coil separately. At iteration $k$, the temporally averaged data is subtracted from corresponding coil data at each time frame, and the subtracted data and temporally averaged data from multi-coils are then inverse Fourier transformed and sensitivity-combined back to image space. This  yields a sequence of aliased images and a temporally averaged sequence (Eq. \ref{temporal_average}). Each frequency-encoding position of the coil-combined images is then processed separately hereafter. The image rows from aliased images or baseline images are gathered and  temporal Fourier transformed along $t$ to yield an $x$-$f$ image, corresponding to $\boldsymbol{\rho}^{k}-\bar{\boldsymbol{\rho}}$ and $\bar{\boldsymbol{\rho}}$ respectively. These signals are then fed as inputs to $xf$-CRNN for $x$-$f$ space reconstruction (Eq. \ref{eq:xf_crnn_rho} and Eq. \ref{xfcnn}). 
After the signal de-aliasing in $x$-$f$ domain, another inverse Fourier transform along $f$ is adopted to transform the estimated $x$-$f$ signal $\boldsymbol{\rho}^{k+1}$ back to dynamic image space for the subsequent weighted coupling with other predictions, as shown in Fig. \ref{fig:architecture}.

\subsubsection{$xt$-CRNN}
Corresponding to the formulation in Eq. \ref{eq:proximal_xyt}, we additionally propose to learn a regulariser in {the spatio-temporal image domain complementary to the combined spatial and temporal frequency domain. Specifically, to effectively exploit the spatio-temporal redundancies in $x$-$y$-$t$ space, we adopt a variation of our previous CRNN-MRI \cite{qin2019convolutional} network for image space de-aliasing which has been shown to be an effective technique in dynamic MRI reconstruction, termed as $xt$-CRNN.} In detail, bidirectional CRNN layers \cite{qin2019convolutional} with recurrence evolving over both temporal and iteration dimensions via hidden-to-hidden connections are employed. This allows us to embed the iterative reconstruction process in a learning setting as well as to propagate information along temporal axis bidirectionally. Similar to the $x$-$f$ space reconstruction, the proposed $xt$-CRNN also learns to reconstruct the combined data from all coils, and learns the residuals of the temporal average baseline $\bar{\mathbf{m}}$ (Eq. \ref{temporal_average}) in spatio-temporal domain with $\mathbf{m}^k-\bar{\mathbf{m}}$ as input to the network. {This can require fewer $k$-$t$ samples for residual encoding and similarly enables the $xt$-CRNN to focus more on the dynamics of the reconstruction. The $x$-$t$ domain and $x$-$f$ domain reconstructions are complementary, which further enables the network to maximally explore cross-domain knowledge for the signal recovery.}
% In particular, to further utilise the spatial-temporal information, we propose to use both the temporally averaged image sequence $\bar{\mathbf{m}}$ together with the reconstruction from the $k$-th iteration $\mathbf{m}^k$ as input to the $im$-CRNN, where $\mathbf{m}^k$ and  $\bar{\mathbf{m}}$ are concatenated along channel axis. This additional information enables the network to better utilise more data for the current frame reconstruction, which can facilitate the training and outperform the single input ($\mathbf{m}^k$) scenario. 
% Previous approaches \cite{eo2018kiki} have shown that exploring cross-domain knowledge is beneficial for MR reconstruction task. Inspired by this, with the aim of exploiting redundancies in complementary domains, here we propose to learn a dynamic MR reconstruction network in both $x$-$f$ and image spaces jointly. In particular, we employ the CRNN model for image space reconstruction due to its effectiveness in exploiting temporal redundancies with a relatively smaller network capacity \cite{qin2019convolutional}. Thus, in each cascade, the proposed $k$-$t$ NEXT consists of a $xf$-CNN and a CRNN block, where it employs all 2D convolutions across spatial and temporal dimensions, in contrast to 3D convolutions used in the baseline method \cite{schlemper2018deep}. This enables the network to be more efficient and effective in learning useful and complementary features in $x$-$f$, spatial and temporal space simultaneously.

\subsubsection{Data consistency layer}
As discussed in Section \ref{Sec:problem-formulation}, Eq. \ref{eq:data_consistency} and Eq. \ref{eq:pDC} give a closed-form {solution with no dense matrix inversion}, so that we can {exactly embed} it as a PI data consistency (pDC) layer in the DNN. To make it concise, we reformulate Eq. \ref{eq:pDC} as:
\begin{equation}
\begin{split}
\boldsymbol{\sigma}_i^{k+1} 
&=\mathbf{F}_\text{s}^\text{H}\left[\Lambda\mathbf{F}_\text{s}\mathbf{S}_i\mathbf{m}^k + (1-\lambda_0)\mathbf{v}_i\right], \\
\Lambda_{jj}  &= \begin{cases}
\lambda_0 & \mathbf{D}_{jj}=1\\
1 & \mathbf{D}_{jj} = 0
\end{cases}
\end{split}
\end{equation}
% &= \text{pDC}(\mathbf{m}^k; \mathbf{S}_i, \mathbf{v}_i, \lambda_0, {\cal D}) \\
where $i \in \left\{ {1,2,...,{n_c}} \right\}$ and $\lambda_0 = {\gamma}/{(\lambda+\gamma)}$. The DC in PI is performed coil-wise and point-wise, {which makes it simple and appealing for implementation in DNNs.} Here $\lambda_0$ is a hyperparameter that allows the adjustment of data fidelity based on the noise level of the acquired measurements. 
\subsubsection{Weighted coupling layer}
Similarly, Eq. \ref{eq:weighted_combination} can be formulated as {a weighted coupling (wCP) layer} in DNNs given estimations from Eq. \ref{eq:proximal_xyf}, Eq. \ref{eq:proximal_xyt} and Eq. \ref{eq:data_consistency}, as represented in the closed-form solution Eq. \ref{eq:WAL}. {The coil sensitivity maps can be normalised to one along coil dimension,} and thus we can simplify Eq. \ref{eq:WAL} as
\begin{equation}
% \footnotesize{
    \mathbf{m}^{k+1} 
    =\alpha_0\mathbf{u}^{k+1}+\beta_0{\mathbf F}_\text{t}^\text{H}\boldsymbol{\rho}^{k+1}+(1-\alpha_0-\beta_0)\sum\limits_{i = 1}^{{n_c}} {\mathbf{S}_i^H} \boldsymbol{\sigma}_i^{k + 1},
    % }
\end{equation}%
% & = \text{WA}({\cal F}_t^{H}\boldsymbol{\rho}^{k+1}, \mathbf{u}^{k+1}, \mathbf{S}_i^H\boldsymbol{\sigma}_i^{k+1}; \alpha_0, \beta_0)
%      \\
in which $\alpha_0 = \frac{\alpha}{\alpha+\beta+\gamma}$ and $\beta_0=\frac{\beta}{\alpha+\beta+\gamma}$ control the weighted coupling of predictions from $x$-$t$ domain and $x$-$f$ domain respectively.
{\subsubsection{CTFNet}
Based on the proposed four modules, our CTFNet can thus be compactly represented as follows:
\begin{subequations}
\label{eq:compact_represent} 
\begin{align}
    \boldsymbol{\rho}^{k+1} &= {\mathbf{F}}_\text{t}\bar{\mathbf{m}}+ xf\text{-CRNN}({\mathbf{F}}_\text{t} \mathbf{m}^k; {\mathbf{F}}_\text{t}\bar{\mathbf{m}}), \label{xfcnn}
\\ 
    \mathbf{u}^{k+1}&=\bar{\mathbf{m}}+xt\text{-CRNN}(\mathbf{m}^k; \bar{\mathbf{m}}), \label{crnn}
    \\
    \boldsymbol{\sigma}_i^{k+1} &= \text{pDC}(\mathbf{m}^k; \mathbf{S}_i, \mathbf{v}_i, \lambda_0, {\mathbf{D}}), \ i \in \left\{ {1,2,...,{n_c}} \right\},\label{pDC}
    \\
    \mathbf{m}^{k+1} & = {\text{wCP}}({\mathbf{F}}_\text{t}^\text{H}\boldsymbol{\rho}^{k+1}, \mathbf{u}^{k+1}, \mathbf{S}_i^\text{H}\boldsymbol{\sigma}_i^{k+1}; \alpha_0, \beta_0).\label{WAL}
\end{align}
\end{subequations}
Here $\bar{\mathbf{m}}$ denotes the temporally averaged sensitivity-combined image of a sequence that is used as the baseline signal, and it can be mathematically expressed as 
\begin{equation}\label{temporal_average}
 \bar{\mathbf{m}}=\sum\nolimits_{i = 1}^{{n_c}} {\mathbf{S}_i^\text{H}} \mathbf{F}_\text{s}^\text{H} \left[\text{max}(\mathbf{I}, \sum\nolimits_{t} \mathbf{D})^{-1}\sum\nolimits_t \mathbf{v}_i\right]_\text{T}, 
\end{equation}
in which max operation is performed element-wise, $\sum\nolimits_{t}$ indicates summation along the temporal dimension, and $\left[ \cdot \right]_\text{T}$ represents the repetition operation along the temporal dimension for T times (the number of frames in a sequence).
Given the proposed framework, our CTFNet can iteratively learn to reconstruct the true images from both spatio-temporal and temporal frequency spaces, so that the spatio-temporal redundancies can be jointly exploited from complementary domains for better reconstructions.}

\section{Methods}
\subsection{Network Architecture and Learning}
\label{network_learning}
The detailed network architecture of the proposed CTFNet is shown in Fig. \ref{fig:architecture}.
{The CTFNet architecture consists of four components, where each component is corresponding to each subequation in Eq. \ref{eq:compact_represent}, respectively. In detail, $xf$-CRNN is composed of 4 layers of CRNN-i and 1 layer of 2D CNN with a residual connection from the baseline estimate.} {The 2D convolutions are applied over the $x$ and $f$ dimensions, and the $y$ dimension is viewed as the batch dimension.} {For the $xt$-CRNN model, a variation of architecture \cite{qin2019convolutional} is employed which consists of 4 layers of BCRNN evolving over both temporal and iteration dimensions, 1 layer of 2D CNN and a residual connection.} {Here the 2D convolutions are applied on the spatial dimensions with recurrence over time dimension in BCRNN, whereas in 2D CNN layer the time dimension is viewed as the batch dimension.} {We used dilated 2D convolutions with kernel size $3 \times 3$ and dilation factor $3 \times 3$ to increase of the receptive field sizes}. {The number of input and output channels of the network was 2, representing the real and imaginary part of the complex-valued data.}

Given the training set $\Omega$ with undersampled data $\mathbf{m}^0$ as input and fully sampled data as target, the network is trained end-to-end by minimising the pixel-wise L1 norm between the reconstructed data and the sensitivity-weighted ground truth data $\mathbf{m}_\text{gt}$:
\begin{equation}
\mathcal{L}\left( \boldsymbol{\theta}  \right){\rm{ = }}\frac{1}{n_\Omega}\sum\limits_{(\mathbf{m}^0, \mathbf{m}_\text{gt}) \in \Omega} {\left\| {{{\mathbf{m}}_\text{gt}} - {{\mathbf{m}}^{n_{it}}}} \right\|_1},
\end{equation}
where $\mathbf{m}^{n_{it}}$ denotes the predicted image at iteration $n_{it}$, i.e., the final output of the proposed network, $\boldsymbol{\theta}$ is the set of network parameters, and ${n_\Omega}$ is the number of training samples. In our setting, we have the iteration step $n^{it}$ set to 5. {{Specifically, all the components in CTFNet including the pDC and wCP layers are needed for training and involved in gradient backpropagation, where the backward pass of the pDC operation can be similarly derived as in \cite{schlemper2018deep} and backward pass for wCP operation with respect to input layers are merely their corresponding coefficients due to the weighted coupling operation.}}
% During the training, the iteration step $n^{it}$ is set to 5, and we used the entire time sequence for training.
For stability of training, values of $\lambda_0$, $\alpha_0$ and $\beta_0$ were all set to 0.1 based on our preliminary works \cite{duan2019vs,hammernik2019sigma}.

{{For training details, gradients were hard-clipped to [-5, 5] to avoid the gradient explosion problem in training recurrent neural networks. The ADAM optimiser was employed with a learning rate of $10^{-4}$. The intensities of input images were normalised by the maximum value of their corresponding undersampled temporal average frame. During training, we extracted training patches along the frequency-encoding direction and used the entire sequence of the data. Networks for different undersampling factors were first trained jointly and then finetuned separately. A plateau of the performance can be observed with $10^5$ backpropagations. Patch extraction and data augmentation were performed on-the-fly on the individual coil images, with random rotation and scaling, and the minibatch size during the training was set to 1.}}

\subsection{Data}
We used two datasets for the experimental evaluations. The first dataset (Dataset A) includes 38 sets of complex-valued multi-slice short-axis cardiac MRI scans acquired on a 1.5T Siemens scanner. 2D bSSFP cine acquisition with retrospective gating and $2\times$ GRAPPA acceleration was performed for 14 healthy subjects and 24 patients {with suspected cardiovascular diseases} for left ventricular coverage. {{In the patient population, we encountered myocarditis, arrhythmogenic right and left ventricular cardiomyopathy, restrictive cardiomyopathy, dilated cardiomyopathy, hypertensive cardiomyopathy, non-ischaemic cardiomyopathy, embolic myocardial infarction and eosinophilic granulomatosis with polyangiitis (EGPA) with cardiac involvement \cite{kustner2020cinenet}.}} The data was acquired with Cartesian sampling and with acquisition parameters including in-plane resolution of $1.9\times1.9$mm, slice thickness of 8mm, {{number of phase-encoding lines of 156, repetition time (TR) of 2.12ms, echo time (TE) of 1.06ms}} and temporal resolution of around 40ms. Images were reconstructed from the $2\times$ acceleration to a fully sampled k-space by GRAPPA. Written informed consent was obtained from all subjects and the study was approved by the local ethics committee (healthy subjects: London Bridge Research Ethics Committee, patients: North of Scotland Research Ethics Committee). In experiments, six slices from each subject that cover the dynamic anatomy were extracted, resulting in a total number of 228 slices for the experiments. Each acquisition in this cohort consists of 25 frames with 30/34/38-channel multi-coil data. The second dataset (Dataset B) used in our experiments consists of 10 fully sampled complex-valued short-axis cardiac cine MRI acquired on a 1.5T Philips scanner. The data were acquired on healthy volunteers following written informed consent under approved research ethics (08/H0711/82). Each scan contains a single slice SSFP acquisition with 30 temporal frames and 32-channel multi-coil raw data. The acquisition has an in-plane resolution of $1.7 \times 1.7$mm, slice thickness of 10mm, {{190 phase-encoding lines, TE=1.66ms, TR=3.32m and an average temporal resolution of 33.70ms}.
}
A variable density incoherent spatiotemporal acquisition (VISTA) sampling scheme \cite{ahmad2015variable} was employed to undersample the k-space data in our experiments, which has been shown to be an effective Cartesian sampling strategy for dynamic data. The scheme is based on a constrained minimisation of Riesz energy on a spatiotemporal grid. It allows uniform coverage of the acquisition domain with regular gaps between samples and guarantees a fully-sampled, time-averaged k-space to facilitate GRAPPA or ESPIRiT kernel estimation. In experiments, we undersampled the data at acceleration rates of 8, 16 and 24, and examples of them are shown in Supporting Information Fig. S1.
Coil sensitivity maps were pre-computed from the fully-sampled, time-averaged k-space center with the ESPIRiT algorithm \cite{uecker2014espirit} by using the BART toolbox \cite{Uecker2015a}.

{\subsection{Experiments}
We firstly performed the comparison study where we compared our CTFNet against other competing approaches on Dataset A with mixed healthy subjects and patients for reconstructions from undersampling rates of 8, 16 and 24. {Here the models were trained on Dataset A with a 2-fold cross-validation, where each fold contained 7 healthy subjects and 12 patients with six slices for each subject.} In the second step, we explored the generalisation potential of the proposed method. {Specifically, we first investigated the robustness of the models when applied to data that were acquired with different scanners and acquisition settings from the training data. We employed models trained on Dataset A and directly tested them on Dataset B. Dataset B differs from Dataset A on the aspects of scanners, acquisition parameters, temporal resolutions, number of acquisition coils and sampling matrix size. In addition, we further investigated the generalisation performance of the proposed method from healthy subjects to patients that were not represented in the training set. In detail, we trained another model with only healthy subjects (14 subjects, 84 slices), and directly tested it on patients in Dataset A. To better understand the proposed method and its performance, an ablation study was also conducted on both datasets to gain more insights on the effects of regularisation in different domains for the dynamic parallel reconstruction problem. Specifically, we investigated and compared between the single domain regularisation ($\mathcal{R}_\text{xt}$ or $\mathcal{R}_\text{xf}$) and the complementary time-frequency domain regularisation.} {{Lastly, a clinical evaluation of the proposed method was performed to investigate the clinical utility of the reconstructed data. In this regard, we measured the left ventricle end-diastolic volume (LVEDV), left ventricle end-systolic volume (LVESV) and left ventricle ejection fraction (LVEF) from the reference standard and the reconstructed accelerated data respectively. The segmentation masks were extracted by using a state-of-the-art DL-based segmentation algorithm \cite{chen2020improving,chen2020realistic}, where the model was previously trained on a different set of cardiac MR data without acceleration.}}

\subsubsection{Evaluation Method}
We compared our proposed approach (CTFNet) with representative MR reconstruction methods, including state-of-the-art CS and low-rank based method $k$-$t$ SLR \cite{lingala2011accelerated}, and two variants of DL methods, dynamic VN \cite{hammernik2019dynamic} and Cascade CNN \cite{schlemper2018deep,schlemper2019data}, which have been substantially enhanced to adapt to dynamic parallel image reconstruction.
Dynamic VN \cite{hammernik2019dynamic} learns the complex spatio-temporal convolutions in contrast to the original VN \cite{hammernik2018learning}, and for strong comparisons with our method, we propose to improve it by incorporating the temporal average baseline as an initialisation.
Similarly, as to Cascade CNN with the D-POCSENSE framework \cite{schlemper2019data} originally designed for static PI, we also refined it to learn the residual of the temporal average, and adjusted it with the same convolutional recurrent architecture as CTFNet to equip it with the ability to exploit spatio-temporal correlations. Thus we term it as CascadeCRNN. The network architecture for CascadeCRNN was the same as $xt$-CRNN and the number of iteration steps $n^{it}$ was set to 5 for all DL methods.
% In addition, to equip it with the ability to exploit spatio-temporal correlations as well as to model the iterative reconstruction process, the Cascade CNN network has also been adjusted with the same convolutional recurrent architecture as the proposed CTFNet, and thus we term it as CascadeCRNN. 
$k$-$t$ SLR formulation has also been extended to be used with multi-coil data based on SENSE model in contrast to its original implementation \cite{lingala2011accelerated}.

Quantitative results were evaluated in terms of normalised mean-squared-error (NMSE) and peak-to-noise-ratio (PSNR) on complex-valued images, as well as structural similarity index (SSIM) and high frequency error norm (HFEN) on magnitude images. These metrics were made to evaluate the reconstruction results with complimentary emphasis. All quantitative results were computed only around cropped dynamic regions for better evaluation. Lower NMSE/HFEN and higher PSNR/SSIM indicate better results. 
% Evaluations on comparison and ablation studies were done via a 2-fold cross-validation on two datasets separately.
}

{{Statistical tests were performed for results of each metric to ensure that the differences between methods were significant.
For multi-model comparisons, we first performed a Friedman test \cite{Friedman1937} to see if there was a significant difference in the population statistics (metric results).
Then if the null hypothesis of the Friedman Test was rejected, we performed one-versus-all one-way Wilcoxon signed-rank test~\cite{Wilcoxon1992} with Bonferroni correction to find out if the results from our model significantly outperformed the others.}}

\subsubsection{Implementation details} 
% The detailed network architecture of the proposed CTFNet is shown and explained in Fig. \ref{fig:architecture}.{\color{blue}$xf$-CRNN is composed of 4 layers of CRNN-i and 1 layer of 2D CNN with a residual connection from the baseline estimate. For the $xt$-CRNN model, a variation of architecture \cite{qin2019convolutional} is employed which consists of 4 layers of BCRNN evolving over both temporal and iteration dimensions, 1 layer of 2D CNN and a residual connection. We used dilated 2D convolutions with kernel size $3 \times 3$ and dilation factor $3 \times 3$. The number of input and output channels of the network was 2, representing the real and imaginary part of the complex-valued data.}Values of $\lambda_0$, $\alpha_0$ and $\beta_0$ were all set to 0.1 based on our preliminary works \cite{duan2019vs,hammernik2019sigma}. 
The CTFNet approach as well as the compared DL methods were all implemented in PyTorch, and trained with the setting as described in Section \ref{network_learning}. {Experiments were performed on a 12GB Nvidia Titan Xp Graphics Processing Unit (GPU).}
% {The network architecture for CascadeCRNN was the same as $xt$-CRNN and the number of iteration steps $n^{it}$ was set to 5 for all methods.} 
% During training, we extracted training patches along the frequency-encoding direction and used the entire sequence of the data. Networks for different undersampling factors were first trained jointly and then finetuned separately, with a total number of $10^5$ backpropagations. Patch extraction and data augmentation were performed on-the-fly on the individual coil images, with random rotation and scaling. 
{For $k$-$t$ SLR, we used the Matlab implementation provided by \cite{lingala2011accelerated} with an extension to multi-coil data. {Experiments were conducted on a 16GB RAM, 3.60GHz Central Processing Unit (CPU).} }

\section{Results}

\subsection{Comparison study}
\begin{figure}[!t]
\centering
{\includegraphics[width=.9\linewidth]{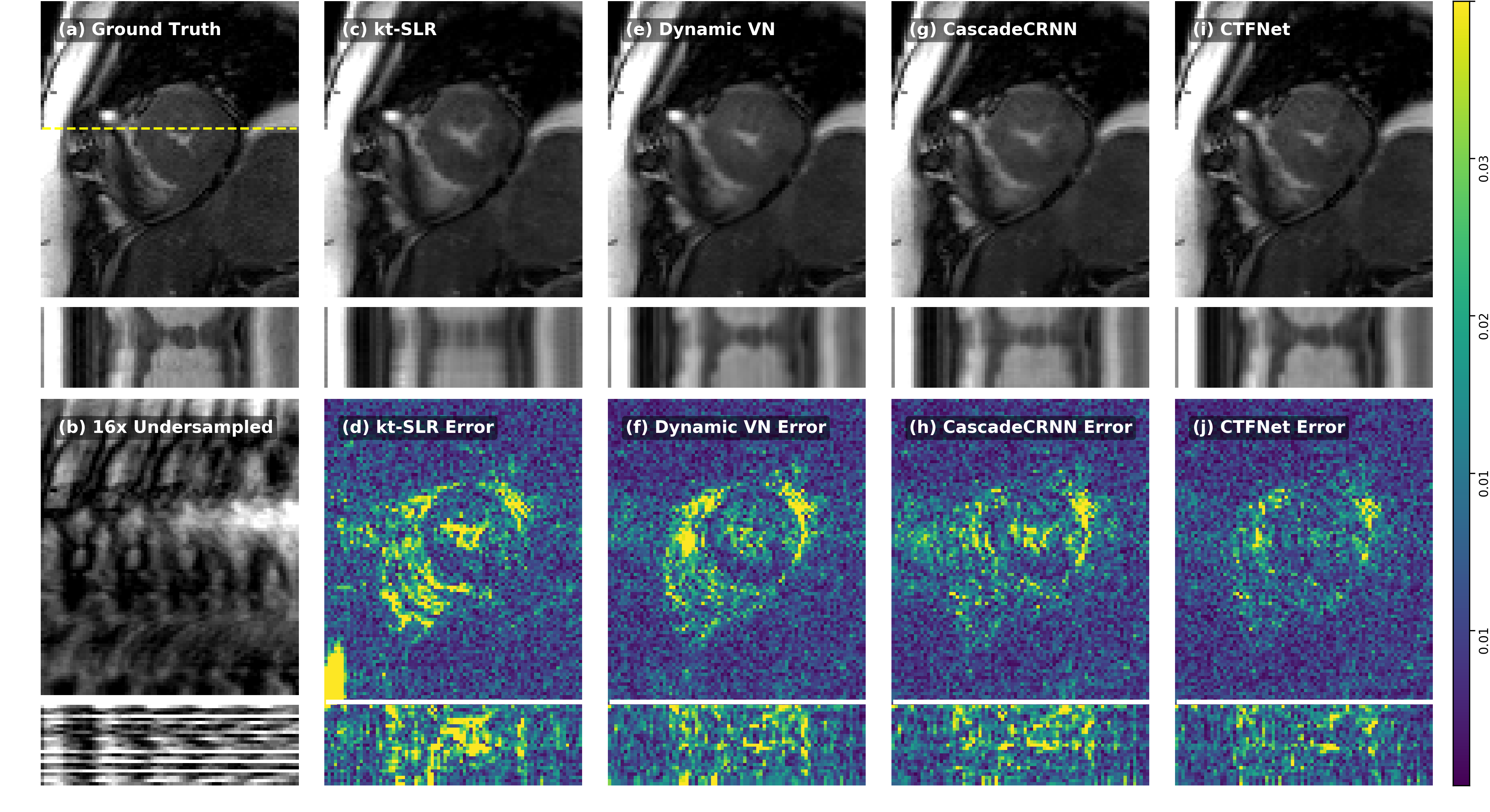}}
\vspace{2mm}
\includegraphics[width=.9\linewidth]{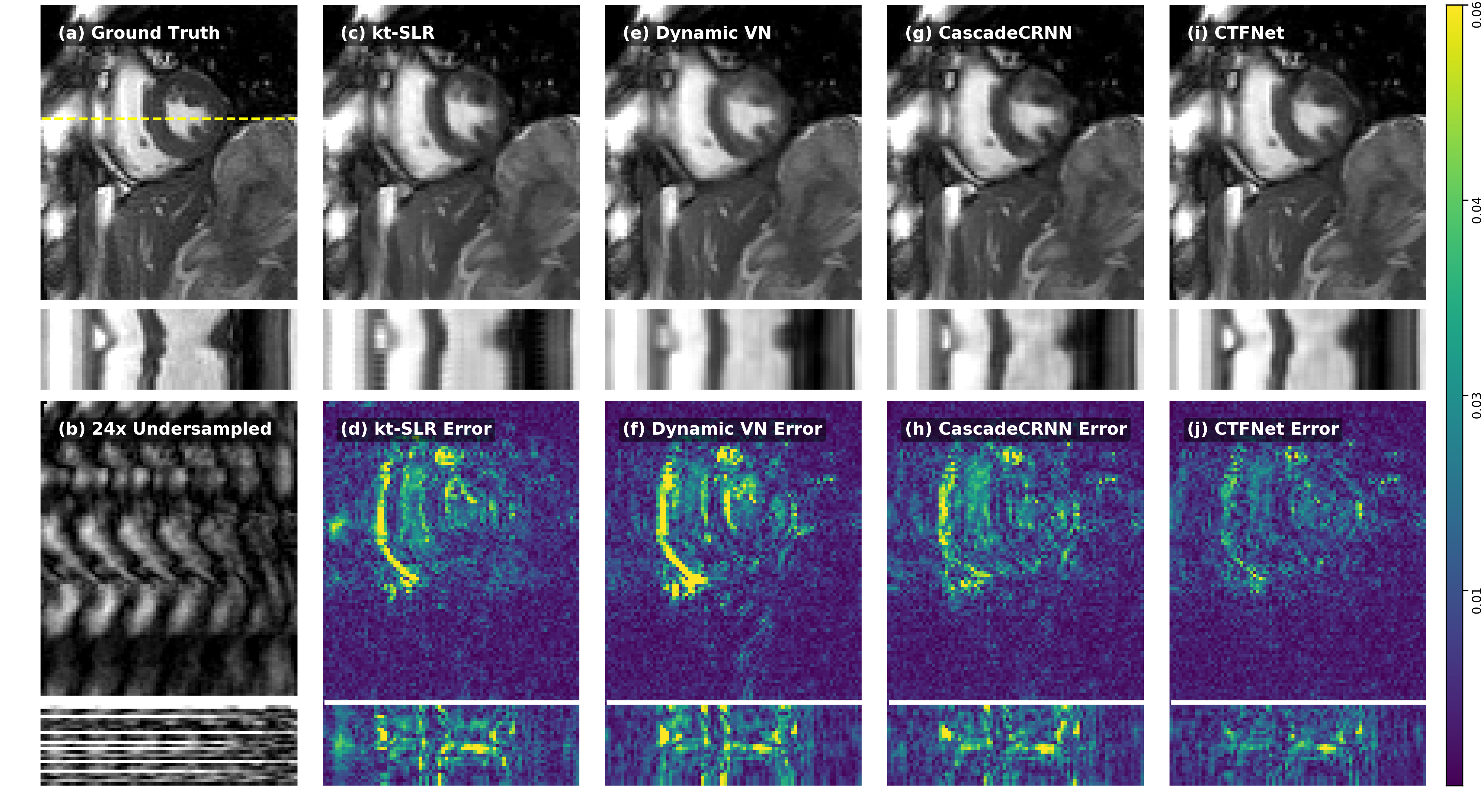}
\caption{Qualitative comparison results of different methods on spatial and temporal dimensions with their error maps. {Results are shown for undersampling rates $16\times$ of a patient (top) and $24\times$ of a healthy subject (bottom) on Dataset A at systolic frames. The scan time for these two acquisitions are 15s and 10s within a single breath-hold respectively. The proposed method can well recover the fine details and preserve the temporal traces, though this gets more challenging on aggressively undersampled data.} An example of the results at diastolic frames is shown in Supporting Information Fig. S2. }
\label{fig:comparison_study}
\end{figure}

Quantitative comparison results of different methods on dynamic multi-coil cardiac data with various high acceleration rates ($8 \times$, $16 \times$ and $24 \times$) are presented in Table \ref{results_2}. {\mdeleted{Here the models were trained on Dataset A with a 2-fold cross-validation, where each fold contained 7 healthy subjects and 12 patients with six slices for each subject.}} The results reported were on the entire 228 2D+t slices on Dataset A. 
% {\st{Since the running time for $k$-$t$ SLR is very slow (over 3 hours for one slice), we only performed $k$-$t$ SLR on mid-ventricle slice from each subject, and the results should be representative.}} 
It can be seen that our proposed CTFNet outperforms $k$-$t$ SLR by a large margin in terms of all these measures at different undersampling rates. {It also offers a much faster ($\sim$1000$\times$) reconstruction speed with 2.8s for the entire sequence of one slice {\mdeleted{(12G TITAN Xp GPU)}} compared with $k$-$t$ SLR with 2444.8s {\mdeleted{(16GB RAM, 3.60GHz CPU)}} for the same reconstruction.} In comparison to other DL-based methods which have been carefully enhanced to incorporate temporal information, our proposed approach can still achieve better performance on all acceleration rates, {with an improvement of around 1dB PSNR and 1.5\% SSIM increase over the most competing method (CascadeCRNN). The performance gap of the improvement is also increasing as acceleration rate increases.} {{All results were statistically significant with $p \ll 10^{-5}$.}} Additionally, we also compared the qualitative results on $16 \times$ {and $24 \times$} undersampled data { (equivalent scan time: 15s and 10s respectively within a single breath-hold)} in Fig. \ref{fig:comparison_study} and Supporting Information Fig. S2, which shows the reconstructed images along both spatial and temporal dimensions as well as their corresponding error maps {on a patient and a healthy subject}. Compared to other competing methods, it can be observed that our proposed model can faithfully recover the images with smaller errors especially around dynamic regions, {and can also produce sharper reconstructions along temporal profiles.}
% The improvement of the proposed method over other competing approaches is also getting more eminent on even higher accelerations ($24 \times$).}

\begin{table*}[!t]
  \centering
  \caption{Comparison results of different methods on Dataset A of dynamic multi-coil cardiac cine MRI with high acceleration rates ({R}). Results (mean (standard deviation)) were computed and compared only around dynamic regions. NMSE is scaled to $10^{-2}$. Best results are shown in bold.}
  \label{results_2}
  {
  \setlength{\tabcolsep}{4.5pt}
  \begin{tabular}{cccccc}  
    % \toprule
    \headrow
 & Metrics & {$k$-$t$ SLR}   & Dynamic VN  & CascadeCRNN &  Proposed    \\
%   \midrule
%   \multicolumn{2}{c}{capacity} & - & 260,866&352,770 & 265,474&374,020\\
%   \midrule
    \hiderowcolors
\multirow{4}{0.5cm}{R=8} &NMSE &0.664 (0.380) &0.529 (0.518) &0.545 (0.516) &\textbf{0.401} (0.314) \\
& PSNR & 40.892 (2.875) & 37.196 (4.786) & 36.945 (4.734) & \textbf{38.051} (4.524) \\
& SSIM & 0.957 (0.023) & 0.970 (0.026) & 0.968 (0.029) & \textbf{0.974} (0.020)\\
& HFEN & 0.138 (0.047)&0.103 (0.076) & 0.110 (0.074) & \textbf{0.087} (0.052)\\
  \midrule
\multirow{4}{0.5cm}{R=16}& NMSE & 1.932 (3.517) & 1.351 (1.012) & 1.253 (1.308) & \textbf{0.947} (0.794)\\
& PSNR &  37.612 (3.136)  & 33.070 (4.648) & 33.372 (4.617) & \textbf{34.384} (4.491)\\
 & SSIM & 0.920 (0.052)&0.936 (0.045) & 0.937 (0.049)& \textbf{0.947} (0.039)\\
 & HFEN & 0.257 (0.154) &0.212 (0.111) &0.194 (0.106) & \textbf{0.166} (0.088)\\
  \midrule
\multirow{4}{0.5cm}{R=24} & NMSE & 2.702 (1.763) & 1.964 (1.734) &1.844 (1.797) & \textbf{1.396} (1.201)\\
& PSNR & 35.222 (3.123) &31.405(4.487)& 31.709 (4.534) &\textbf{32.713} (4.399)\\
 & SSIM & 0.895 (0.052)&0.914 (0.055) & 0.914 (0.060) & \textbf{0.929} (0.049)\\
 & HFEN &0.309 (0.107) &0.270 (0.124) & 0.251 (0.123)& \textbf{0.215} (0.104)\\
    \bottomrule
  \end{tabular}}
\end{table*}

\subsection{Generalisation study}
\begin{figure}[!t]
\centering
\includegraphics[width=.9\linewidth]{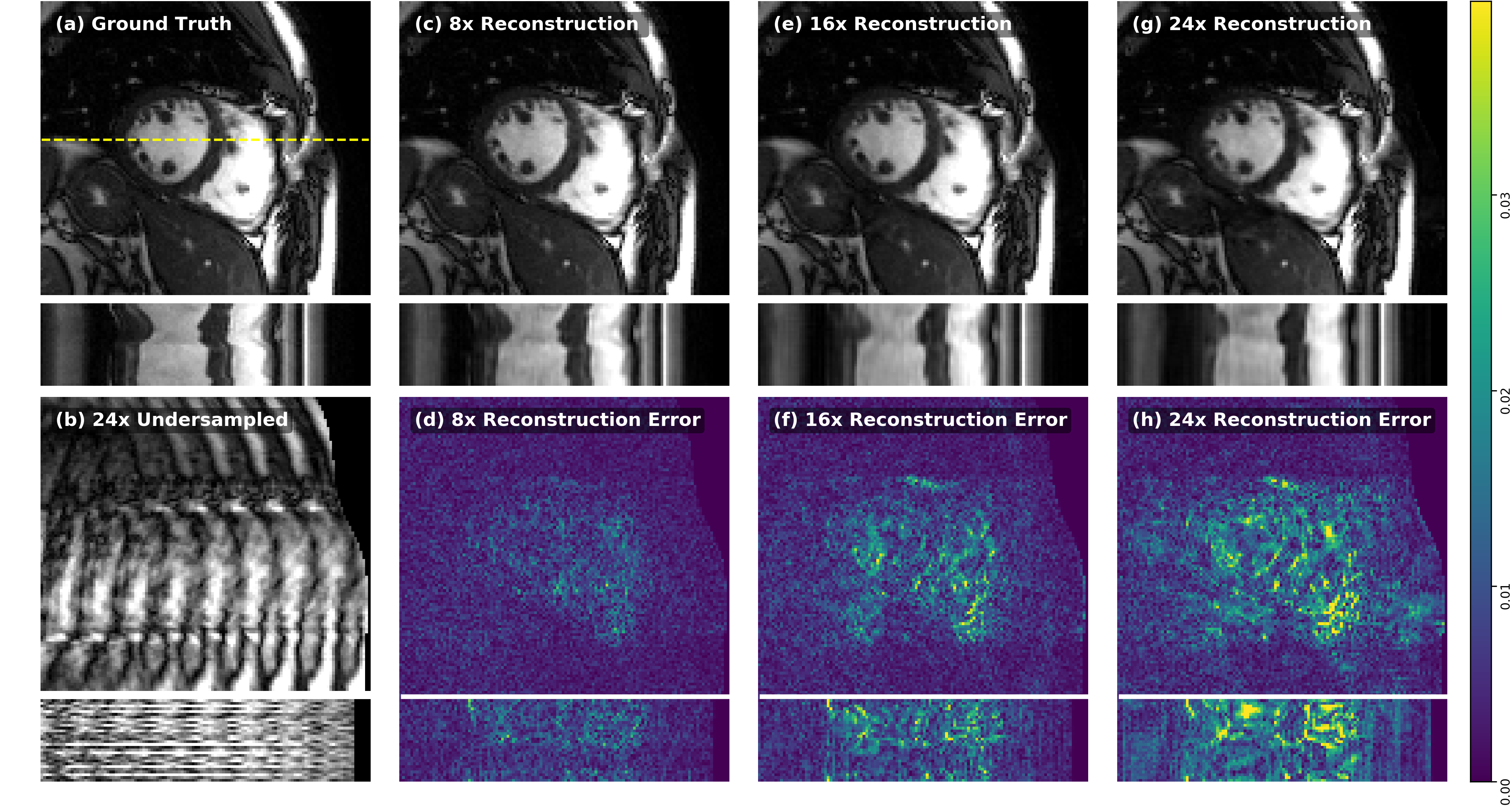}
\caption{{Generalisation reconstructions of the proposed method on the unseen domain Dataset B along spatial and temporal dimensions} with various acceleration rates as well as their error maps. (a) Fully sampled image (b) Example of undersampling image with {R=24} (c) (d) Reconstruction from {R=8} (e) (f) Reconstruction from {R=16} (g) (h) Reconstruction from {R=24}. {The proposed method can well reconstruct the images with good preservation of temporal trace on various undersampling rates. Though reconstruction is more challenging as {R} increases, the reconstructed results can still be useful.}}
\label{fig:xt_comparison}
\end{figure}

{In this study, we explored the generalisation potential of the proposed method.} {\mdeleted{We first investigated the robustness of the models when applied to data that were acquired with different scanners and acquisition settings from the training data. Specifically, we employed models trained on Dataset A and directly tested them on Dataset B. Dataset B differs from Dataset A on the aspects of scanners, acquisition parameters, temporal resolutions, number of acquisition coils and sampling matrix size.}} The generalisation test results of different DL models from Dataset A to Dataset B are presented. {{Particularly, to better understand how the methods perform on the key frames, here we show their comparisons on the end-systolic (ES) frame (Table \ref{results_generalisation_scanner}) and end-diastolic (ED) frame respectively (Supporting Information Table S1).}} It can be seen that the proposed method achieves high performance on the unseen test dataset and also consistently outperforms against other competing methods, indicating its capability in effectively learning the inverse dynamic reconstruction problem. {{All results on both ES and ED frames were also statistically significant with $p<0.005$.}}
% When comparing results from Table \ref{tab:results_ablation} with Table \ref{results_generalisation} on AF $8\times$ of dataset B, it can be seen that there is a slight decrease of performance on the generalisation results, which is possibly due to the domain shift of the data.
% This is due to the fact more training data is available in dataset A than dataset B (over $20 \times$) when training the models. 
Besides, we also visualised the generalisation results of Dataset B under different acceleration rates, as presented in Fig. \ref{fig:xt_comparison} and Supporting Information Fig. S3. {It can be observed that our approach can recover the fine details and the temporal traces of the image very well on data from unseen domain even with extreme undersampling rate ($24 \times$)}, though it is anticipated that the reconstruction gets more challenging as acceleration rate increases. 
% Similarly, visualisation of $x$-$f$ reconstruction on Dataset B is also shown in Fig. S2, where it displays the reconstructed $x$-$f$ images and their error maps with various undersampling rates. 
% \st{An example of the undersampled $x$-$f$ signal with AF $16\times$ is shown in Fig. S2 (b). It can be seen that the aliasing artefacts were largely removed and the undersampled data were recovered to approximate the ground truth signals with the proposed method.} 
% Both results indicate that our approach is able to recover the signals on data from unseen domain even with extreme undersampling, despite higher errors around dynamic regions than reconstructions from less aggressively undersampled data.

\begin{table*}[!t]
  \centering
  \caption{Generalisation results of different DL methods trained on Dataset A and deployed to Dataset B for different acceleration rates. Results (mean (standard deviation)) were computed and compared only around dynamic regions on ES frame. NMSE is scaled to $10^{-2}$. Best results are shown in bold.}
  \label{results_generalisation_scanner}
  {
  \setlength{\tabcolsep}{4.5pt}
  \begin{tabular}{ccccc}  
    % \toprule
    \headrow
 & Metrics    & Dynamic VN  & CascadeCRNN &  Proposed     \\
%   \midrule
  \hiderowcolors
\multirow{4}{0.5cm}{R=8} & NMSE & 1.596 (0.639) &1.282 (0.492) & \textbf{1.124} (0.390)\\& PSNR & 31.329 (1.674) & 32.236 (2.010) & \textbf{32.763} (1.921) \\
& SSIM & 0.932 (0.019) & 0.943 (0.015) & \textbf{0.948} (0.013)\\
& HFEN & 0.176 (0.036) & 0.160 (0.034) & \textbf{0.147} (0.025)\\
  \midrule
\multirow{4}{0.5cm}{R=16} & NMSE & 3.196 (1.356) & 2.465 (1.005) & \textbf{2.029} (0.689)\\& PSNR   &28.799 (2.186) & 29.905 (2.367)& \textbf{30.647} (2.141)\\
 & SSIM &0.896 (0.026) & 0.905 (0.028)& \textbf{0.919} (0.019)\\
 & HFEN & 0.309 (0.090)&0.241 (0.065) & \textbf{0.214} (0.049)\\
  \midrule
    \multirow{4}{0.5cm}{R=24} & NMSE & 4.837 (1.837)&3.966 (1.373) &\textbf{3.342} (1.064) \\& PSNR &26.873 (1.714) & 27.737 (1.957) & \textbf{28.438} (1.741)\\
 & SSIM & 0.857 (0.027)& 0.867 (0.030) & \textbf{0.887} (0.024)\\
 & HFEN & 0.414 (0.088)& 0.343 (0.066) & \textbf{0.312} (0.061)\\
    \bottomrule
  \end{tabular}}
\end{table*}

\begin{figure}[!t]
\centering
{\includegraphics[width=.65\linewidth]{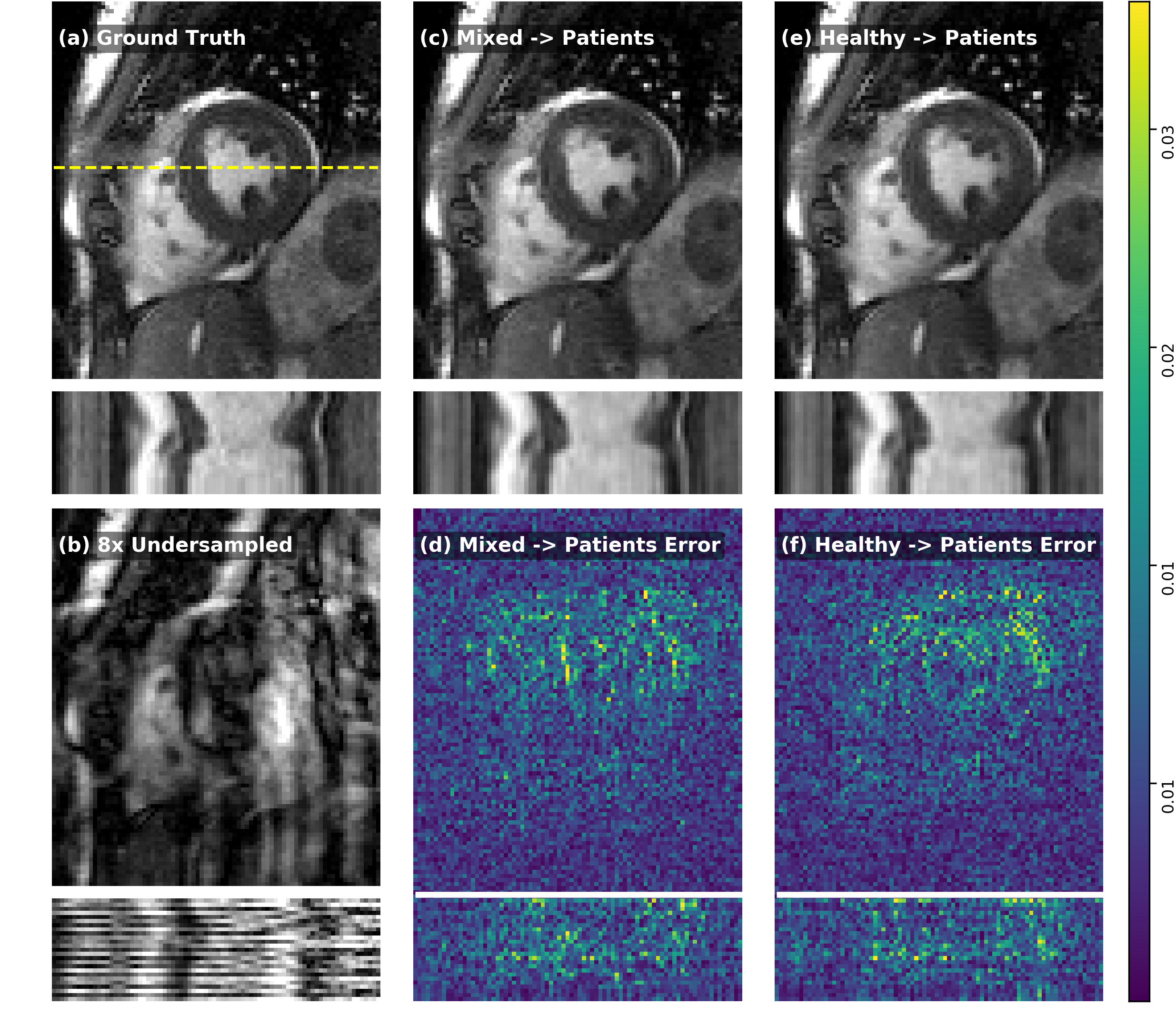}}
\caption{{Comparison of the proposed method between mixed training results (from mixed healthy subjects/patients to patients) and generalisation results (from healthy subjects to patients). Results shown are on one patient with hypertensive cardiomyopathy in Dataset A on {R=8}.} The generalisation result is almost as well as the one from standard mixed training.}
\label{fig:patients_comparison}
\end{figure}

{{In addition, we further investigated the generalisation performance of the proposed method from healthy subjects to patients that were not represented in the training set. In detail, we trained another model with only healthy subjects (14 subjects, 84 slices), and directly tested it on patients in Dataset A.}} In addition, the generalisation results from healthy subjects to patients were compared with models trained with mixed healthy subjects and patients (19 subjects, 114 slices), as shown in Table \ref{tab:results_generalisation}. Though the pathological conditions were not included in the training data, the generalisation results from healthy data to patients were very competitive to the mixed training models with an average of only 0.2dB PSNR and 0.2\% SSIM drop of performance. This can also be observed from the qualitative comparison as shown in Fig. \ref{fig:patients_comparison}, where only subtle differences can be detected from these two training settings.

\begin{table*}[!t]
  \centering
  \caption{{Generalisation results of the proposed method trained on healthy subjects only (84 slices) and tested on patients in Dataset A for different acceleration rates. Results (mean (standard deviation)) were computed only around dynamic regions and compared with models trained with mixed healthy subjects and patients (114 slices) also in Dataset A. NMSE is scaled to $10^{-2}$. Better results are shown in bold.}}
  \label{tab:results_generalisation}
  {
  \setlength{\tabcolsep}{4.5pt}
  \begin{tabular}{cccc}  
    % \toprule
    \headrow
& Metrics    & Mixed (114) $\rightarrow$ patients & healthy (84) $\rightarrow$ patients      \\
%   \midrule
  \hiderowcolors
\multirow{4}{0.5cm}{R=8} & NMSE  &\textbf{0.393} (0.317) & {0.421} (0.366)\\
& PSNR  & \textbf{37.430} (4.552) & {37.275} (4.642) \\
& SSIM  & \textbf{0.971} (0.023) & {0.969} (0.026)\\
& HFEN  & \textbf{0.094} (0.057) & {0.096} (0.066)\\
  \midrule
% \multirow{4}{0.5cm}{$16 \times$} & NMSE  & 0.939 (0.842) & {0.981} (0.849)\\& PSNR    & 40.563 (3.665)& {40.379} (3.700)\\
%  & SSIM  & 0.939 (0.046)& {0.938} (0.046)\\
%  & HFEN &0.180 (0.098) & {0.183} (0.099)\\
\multirow{4}{0.5cm}{R=16} & NMSE  & \textbf{0.909} (0.795) & {0.981} (0.849)\\& PSNR    & \textbf{33.825} (4.482)& {33.537} (4.563)\\
 & SSIM  & \textbf{0.941} (0.044)& {0.938} (0.046)\\
 & HFEN &\textbf{0.176} (0.095) & {0.183} (0.099)\\
  \midrule
% \multirow{4}{0.5cm}{$24 \times$} & NMSE &1.347 (1.178) &{1.353} (1.091) \\
% & PSNR  & 38.902 (3.610) & {38.855} (3.615)\\
%  & SSIM & 0.919 (0.055) & {0.919} (0.055)\\
%  & HFEN & 0.228 (0.112) & {0.232} (0.113)\\
\multirow{4}{0.5cm}{R=24} & NMSE &\textbf{1.325} (1.184) &{1.353} (1.091) \\
& PSNR  & \textbf{32.177} (4.330) & {32.013} (4.341)\\
 & SSIM & \textbf{0.921} (0.055) & {0.919} (0.055)\\
 & HFEN & \textbf{0.224} (0.112) & {0.232} (0.113)\\
    \bottomrule
  \end{tabular}}
\end{table*}

\subsection{Ablation study}
\begin{figure}[!t]
\centering
\includegraphics[width=0.9\linewidth]{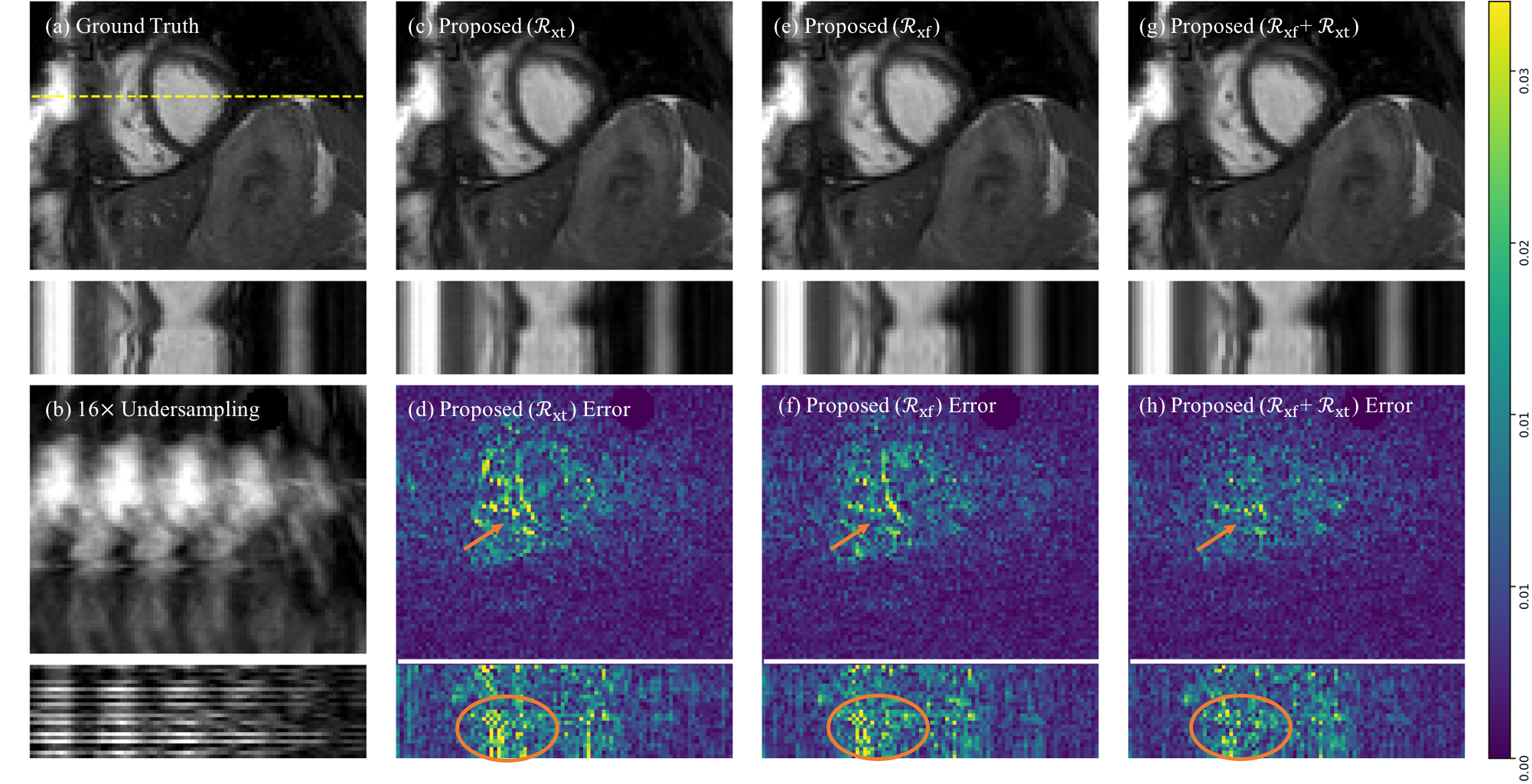}
\caption{Qualitative comparisons of the ablated different domain reconstructions on spatial and temporal dimensions with their error maps. Results are shown for {{R=16} (scan time 15s) on Dataset A. Highlighted regions indicate improvement of the complementary time-frequency domain reconstruction.}}
\label{fig:ablation_study}
\end{figure}

{{In addition, we performed the ablation study to gain more insights on the effects of different regularisations for the problem. Particularly, we investigated on the effects of different regularisations ($\mathcal{R}_\text{xt}$ and $\mathcal{R}_\text{xf}$) on the dynamic parallel reconstruction problem. Specifically,}} The ablation study compared results from the spatio-temporal image space reconstruction (Proposed ($R_\text{xt}$)), the combined temporal Fourier and spatial space reconstruction (Proposed ($R_\text{xf}$)) as well as the complementary time-frequency domain reconstruction (Proposed ($R_\text{xf}+R_\text{xt}$)). All these ablated approaches with varying domain regularisations were conducted under the same variable splitting framework as in Section \ref{Sec:vs-next}, where for the single domain reconstruction, only the corresponding domain network was used. The quantitative comparison results of the ablation study are shown in Table \ref{tab:results_ablation}, where reconstruction models were trained on data with {R=8} from datasets A and B respectively. {A qualitative result is also given in Fig. \ref{fig:ablation_study} on data with {R=16}.}
% It can be seen that all these variants of the proposed method can achieve very good results. In particular, it can be observed that for the dynamic MRI, $x$-$f$ domain reconstruction (Proposed ($R_{xf}$)) achieves better performance both quantitatively and qualitatively than the $x$-$t$ space reconstruction (Proposed ($R_{xt}$)) with even a smaller number of network parameters, which implies that exploiting spatio-temporal correlations in combined spatial and temporal frequency domain is more effective for the specific application. In addition, by combining these two regularisation terms, the proposed CTFNet (Proposed ($R_{xf}+R_{xt}$)) can attain a further improvement as shown in Table \ref{tab:results_ablation} and Fig. \ref{fig:ablation_study} with fewer reconstruction errors, indicating the benefits of jointly exploiting spatio-temporal redundancies from complementary domains.
\subsection{Clinical Evaluation}
{{Our experiments indicate a good reconstruction quality of the proposed method with respect to the reference standard, however, the question of the clinical validity of such reconstructions remains open. 
% In this regard, we measured the left ventricle end-diastolic volume (LVEDV), left ventricle end-systolic volume (LVESV) and left ventricle ejection fraction (LVEF) from the reference standard and the reconstructed accelerated data respectively. The segmentation was performed by using a robust automatic DL-based segmentation tool, where the model was previously trained on a different set of cardiac MR data without acceleration. 
To understand the translational utility, we performed the left ventricular function assessment (LVEDV, LVESV and LVEF) on our reconstructed data in Dataset A.
% Specifically, we compared the clinical variables (LVEDV, LVESV and LVEF) derived from the reconstructions of the proposed method with that from the reference standard on Dataset A.
Note that the DL-based segmentation tool was trained on another group of cardiac MR data of healthy subjects without acceleration (UK Biobank data \cite{petersen2017reference}) and DL-based approaches are also known to be sensitive to variations of image distributions and perturbations, so there were some inevitable failure cases on our test data. Therefore, we performed manual quality control of the segmentation in our cases, where we have excluded severe failure cases from the cohort. The resulting dataset includes 12 healthy subjects and 13 patients, and we show their comparisons with the reference standard in Bland-Altman plots in Supporting Information Fig. S4 for all acceleration rates.
It can be observed that our reconstructions have achieved reasonably good results on all three measures, and on average a bias for LVEDV, LVESV and LVEF of -0.04ml, 1.15ml and -0.97\% was observed with all observations lying inside the 96\% confidence interval of $\pm 2.21\%$, $\pm 2.53$ and $\pm 2.36\%$ on $8\times$ accelerated data. For $16 \times$ and $24\times$ accelerated data, an average bias of -1.81ml, 1.98ml and -2.32\% for LVEDV, LVESV and LVEF was observed with a slightly higher variance on $24\times$ accelerated data than on $16\times$ accelerated data. Both the biases and variances of these measurements are within an acceptable range, indicating that our reconstruction results can potentially have the clinical benefit.
}}

\begin{table*}[!t]
  \centering
  \caption{Ablation study of effects of different regularisations on dynamic cardiac cine MRI reconstruction. Experiments were performed on two different datasets (A and B) with undersampling rate $8\times$. NMSE is scaled to $10^{-2}$. Results are presented in mean (standard deviation). Best results are indicated in bold.}
  \label{tab:results_ablation}
%   \scalebox{0.95}{
  \setlength{\tabcolsep}{4.5pt}
  \begin{tabular}{ccccccc}  
  \headrow
    % \toprule
  \multicolumn{2}{c}{Method} &  Proposed ($\mathcal{R}_\text{xt}$)  & Proposed ($\mathcal{R}_\text{xf}$) &  Proposed ($\mathcal{R}_\text{xf}+\mathcal{R}_\text{xt}$)    \\
%   \midrule
  \multicolumn{2}{c}{\# params} &408,578 & 260,866 & 669,444\\
  \hiderowcolors
     \midrule
\multirow{4}{0.5cm}{A} & NMSE & 0.528 (0.454) & 0.462 (0.407) & \textbf{0.401} (0.314)\\
& PSNR  & 36.932 (4.585) & 37.580 (4.668) & \textbf{38.051} (4.524)\\
& SSIM & 0.969 (0.026)&0.970 (0.026) & \textbf{0.974} (0.020)\\
& HFEN &0.107 (0.068) & 0.096 (0.064) & \textbf{0.087} (0.052)\\
\midrule
\multirow{4}{0.5cm}{B}  & NMSE & 0.906 (0.288)&0.852 (0.274) & \textbf{0.723} (0.197)\\
& PSNR  & 33.523 (1.989) & 33.796 (1.995) & \textbf{34.455} (1.886)\\
& SSIM & 0.956 (0.010)&0.958 (0.011) & \textbf{0.961} (0.009)\\
& HFEN &0.126 (0.027) & 0.115 (0.020) & \textbf{0.105} (0.019)\\

    \bottomrule
  \end{tabular}
\end{table*}

\section{Discussion}
In this work, we have demonstrated that the proposed method is capable of recovering high quality images from highly undersampled dynamic multi-coil data. Different from existing DL-based approaches, we incorporated {the combined spatial and temporal frequency domain} regularisation into the formulation of the dynamic parallel MRI reconstruction problem and exploited spatio-temporal redundancies from both $x$-$t$ and $x$-$f$ spaces with DNNs. Compared with {{spatio-temporal image ($x$-$t$)}} domain reconstruction {(Proposed ($R_\text{xt}$), Table \ref{tab:results_ablation})}, the proposed $x$-$f$ space reconstruction (Proposed ($R_\text{xf}$)) has shown to be more effective in exploiting the spatio-temporal correlations, with higher reconstruction accuracy and a smaller number of network parameters {(Table \ref{tab:results_ablation})}. This is mainly due to the inherent nature of the periodic dynamic cardiac MRI data itself, where strong correlations exist in k-space and time and signal in temporal Fourier space is sparse. This has been represented in many traditional CS-based methods, and here our results have demonstrated that {the learned implicit DNN-prior in the temporal Fourier domain can further increase the acceleration capability and achieve even better performance.} In addition, combination of {time-frequency cross-domain} knowledge {(Proposed ($R_\text{xf}+R_\text{xt}$), Table \ref{tab:results_ablation} and Fig. \ref{fig:ablation_study})} further enhances the reconstruction capability of the proposed method with better reconstruction quality. {In comparison to three fully independent constraints which regularise the spatial, temporal and temporal-frequency dimensions independently, our approach offers the ability to efficiently exploit correlations in the joint spatial and temporal/temporal-frequency space, which enables the estimation of missing data to be based on other available points within its spatial vicinity at neighboring time points or temporal-frequency. Additionally, the use of two constraints is also more efficient than three independent constraints, as adding one extra constraint will lead to another subproblem for minimisation, which thereby will increase complexity in network architecture design.} The improved performance of CTFNet over other competing methods indicates that learning jointly from both spatio-temporal and temporal frequency domains can capture complementary useful information that can be effectively utilised by the proposed framework.
% {which also explains the superior performance of CTFNet over other competing methods}. 

Furthermore, the proposed CTFNet builds on a multi-variable minimisation problem and embeds it into an efficient DL framework. The employed variable splitting technique effectively decouples data regularisation terms on various domains from the data fidelity term, which enables the natural derivation of pDC layer {and wCP layer} in PI with closed-form point-wise solutions.
% This makes the framework computationally efficient {and also enables it simple and appealing for implementation in deep neural networks.}
Though the derived pDC layer shares similar form as the one proposed in D-POCSENSE \cite{schlemper2019data} which is a simple extension from single-coil application \cite{schlemper2018deep}, our solution (pDC with wCP layers) for the multi-coil setting has the mathematical support based on variable splitting and alternating minimisation, and thus reasons the particular formulation and structure of our network. {In contrast to \cite{aggarwal2018modl,biswas2019dynamic} where data fidelity step is solved via conjugate gradient algorithm due to the difficult matrix inversion in their DC terms, our CTFNet offers a much simpler and more efficient solution with exact steps and avoids iterative gradient updates, allowing for faster reconstruction speed and easier embedding into DNNs.}
% In addition, in comparison to VN whose regularisation term is defined via a set of explicit learnable linear filter kernels and activation functions, our proposed CTFNet implicitly learns the data regularisation by leveraging the DNNs with a recurrent structure.
% {\st{When comparing between different frameworks with only image domain regularisation, i.e. Proposed-$\mathcal{R}_s$
% in Table 4 in comparison to CascadeCRNN and Dynamic VN in Table 1 our proposed method (Proposed-$\mathcal{R}_s$) can achieve competitive performance against these competing methods that have been well modified to better exploit spatio-temporal redundancies.}} 
Besides, our approach also offers the flexibility of incorporating additional regularisation terms in the framework, whereas this will not be very straightforward for the other approaches. 

{It is worth mentioning that with the recurrent design of the architecture, it is natural that the parameters for each iteration are shared, whereas this is not the case for other approaches such as VNs \cite{hammernik2018learning}. Similar to other DL-based unrolled network architecture, our proposed approach is a learning-based method which is designed to achieve optimal performance with the pre-set fixed number of iterations during the training, though the recurrent design enables it to be employed for arbitrary number of iterations at inference time. In our work, we determined heuristically that 5 iterations represents a compromise between GPU memory requirements and performance. Therefore, we fixed $n_{it}=5$ to train our network, and thus it is expected that the optimal inference performance also comes from $n_{it}=5$. If $n_{it}$ is increased correspondingly at training stage, we can expect that the performance of the network will be further improved and converge after a few iterations. However, due to the multi-coil setting and the limitation of GPU memory, we were not able to train the model with more iterations. More efficient training schemes will be investigated in the future to improve the performance of the proposed method.} {In addition, with respect to the hyper-parameters, our DL-based method appears to be more robust at test time while CS-based approaches may require case-level tuning for the optimal results.
% note that they can also be simultaneously learnt during the training process. However, for the reason of stable training, here we choose to set these parameters fixed to certain values based on preliminary experiments \cite{duan2019vs,hammernik2019sigma}, where we have also found that the reconstruction performance is insensitive to these hyper-parameter values due to the learning scheme. In comparison to CS-based approaches which heavily rely on application-specific selection and tuning of hyper-parameters in the reconstruction phase, the choice of hyper-parameters in our method is only needed once in the training phase and can then be fixed and directly applied to the reconstruction stage. Therefore, our approach is more robust and efficient compared to CS-based methods in terms of the choice of hyper-parameters. 
}
% {\st{In our work, we also propose to learn the residual of a temporally averaged frame, and we have found that this can improve the reconstruction performance compared with the standard residual learning as from our preliminary research. This is mainly due to the fact that strong redundancy exists along temporal dimension especially in the background regions, and thus learning the residual of a temporal average reduces the efforts in background learning and enables the network to focus more on reconstructing the dynamic regions.}} 

{In our work, we also propose to learn the residual of a temporally averaged frame, which is not necessarily required to be a fully-sampled image. This can be seen from our undersampling masks of R=24 (Supporting Information Fig. S\ref{sfig:masks}), where the averaged k-space is not fully-sampled but good reconstruction results can still be achieved. Besides, VISTA sampling pattern was designed in a way to spread the sampled lines out as much as possible to high frequencies, though it also appears to omit some of the highest k-space locations which is a common phenomenon for high accelerations of undersampling strategies. VISTA has been shown to be able to generate more consistent results with superior noise–sharpness trade-off compared to other commonly employed sampling patterns [48], and thus it is also expected to benefit the DL-based reconstruction. The fundamental acceleration limit for which the undersampling does not affect the effective reconstructed resolution was not investigated in [48], and this is also not the focus in our work. Nevertheless, the proposed approach is a learning-based method, i.e. the network is able to reconstruct the missing high frequency samples from the trained information and by that recover reasonably good reconstructions. Additionally, the impact of motion would also be crucial for understanding the robustness of the model. Theoretically, since the model allows high acceleration rates, it is likely that the model can accommodate even if some data are rejected or omitted. However, it would require comprehensive evaluation to study various degrees and types of motion and artifacts, and this will be an important direction of future work.}

Moreover, the proposed method can generalise well to unseen cardiac MR data with different acquisition parameters {and with pathology that were not seen in the training set}. The method can achieve satisfactory performance on these scenarios even with highly aggressive undersampling strategies, which indicates that the proposed method is robust to unseen and unusual image features or temporal behaviours present in our currently used dataset. 
% This shows promising results for deploying DL models for clinical practice, nevertheless, more validations on this aspect including radiologists' discretion are still needed for its practical use.
{{In addition, our clinical evaluation of the reconstructed images shows acceptable biases and variances on the left ventricular function assessment, indicating the great potential to deploy DL models for clinical practice. Also to note that the differences between the reference and reconstructed data could also be accumulated from errors induced by the segmentation performances especially on accelerated data, thus this shall also be taken into consideration when comparing them with the reference standard. However, the correction and improvement of the segmentation on accelerated data are out of scope of this work, and should be considered as an important avenue for future research.}}

Particularly, {by exploiting spatio-temporal redundancies in the proposed DL framework, our approach can outperform the state-of-the-art CS and DL-based methods and can further push the acceleration capability with fast reconstruction speed for the dynamic parallel MR imaging}. In our work, Dataset A was a multi-breath-hold acquisition of 8 consecutive breath-holds with 15s for each ($2\times$ GRAPPA accelerated). {{Hence, an acceleration rate of 16 or higher could result in the possibility of achieving the same acquisition in a single breath-hold, but this may also require further investigation with respect to transient state imaging, achievable spatial coverage and/or temporal resolution for single breath-hold cine imaging.}} Despite this being a retrospective undersampling study, our results indicate a great potential in facilitating fast single-breath-hold clinical 2D cardiac cine imaging.

For the future work, we will explore the dynamic parallel image reconstruction with other types of undersampling strategies, such as radial sampling which is also commonly used in acceleration of 2D cardiac MR imaging in practice. In addition, we could also consider incorporating some other regularisation terms into the framework, such as regularisation on some other transform domains, to exploit the data redundancy for effective reconstruction. Besides, generalisation capability of the model can be further validated on more data from different domains and with various acquisition parameters {and pathologies}, {{as well as on more aspects such as generalisation to different orientations (e.g. chamber views)} to investigate its potential application for clinical use.
}

\section{Conclusion}
In this paper, we have proposed a novel DL-based approach, {termed CTFNet}, for highly undersampled dynamic parallel MR image reconstruction. The proposed method exploits spatio-temporal correlations in both the combined spatial and temporal frequency domain and the spatio-temporal image domain based on a variable splitting and alternating minimisation formulation. 
% Our approach is based on a variable splitting and alternating minimisation formulation, where an $xf$-CRNN is proposed to reconstruct the true signals from aliased signals in $x$-$f$ space and an $xt$-CRNN is employed to recover the signals in $x$-$t$ space. {Closed-form data consistency and weighted coupling terms have also been derived, which are exactly and explicitly embedded as layers in our CTFNet.} 
The network is able to learn to iteratively reconstruct the images by jointly and effectively exploiting information from the complementary time-frequency domains. Our proposed CTFNet outperforms state-of-the-art dynamic MR reconstruction methods in terms of both quantitative and qualitative performance, {with excellent recovery of fine details and preservation of temporal traces. It also enables increased accelerations of data acquisition with favorable generalisation ability, which is promising for realising single-breath-hold clinical 2D cardiac cine MR imaging.}

\section*{Acknowledgements}
This work was supported by EPSRC programme grant SmartHeart (EP/P001009/1). We thank Chen Chen for providing us with the deep learning based segmentation algorithm.

\section*{Data Availability Statement}
{To support the findings of our manuscript, we make our source code available at \url{https://github.com/cq615/kt-Dynamic-MRI-Reconstruction}.
}

% \begin{appendix}
% \input{appendix}
% \end{appendix}
% \section*{conflict of interest}
% You may be asked to provide a conflict of interest statement during the submission process. Please check the journal's author guidelines for details on what to include in this section. Please ensure you liaise with all co-authors to confirm agreement with the final statement.

%\printendnotes

% Submissions are not required to reflect the precise reference formatting of the journal (use of italics, bold etc.), however it is important that all key elements of each reference are included.
\bibliography{main}

% \begin{biography}[example-image-1x1]{A.~One}
% Please check with the journal's author guidelines whether author biographies are required. They are usually only included for review-type articles, and typically require photos and brief biographies (up to 75 words) for each author.
% \bigskip
% \bigskip
% \end{biography}

% \graphicalabstract{example-image-1x1}{Please check the journal's author guildines for whether a graphical abstract, key points, new findings, or other items are required for display in the Table of Contents.}

\clearpage
\section*{SUPPORTING INFORMATION}

\begin{stable}
  \centering
  \caption{Generalisation results of different DL methods trained on Dataset A and deployed to Dataset B for different acceleration rates. Results (mean (standard deviation)) were computed and compared only around dynamic regions on the ED frame. NMSE is scaled to $10^{-2}$. Best results are shown in bold.}
  \label{results_generalisation}
  {
  \setlength{\tabcolsep}{4.5pt}
  \begin{tabular}{ccccc}  
    % \toprule
    \headrow
 & Metrics    & Dynamic VN  & CascadeCRNN &  Proposed     \\
%   \midrule
  \hiderowcolors
\multirow{4}{0.5cm}{R=8} & NMSE & 1.356 (0.762) &1.057 (0.311) & \textbf{0.765} (0.236)\\& PSNR & 32.053 (2.343) & 32.855 (2.100) & \textbf{34.266} (2.104) \\
& SSIM & 0.939 (0.018) & 0.948 (0.014) & \textbf{0.961} (0.010)\\
& HFEN & 0.171 (0.141) & 0.154 (0.026) & \textbf{0.125} (0.021)\\
  \midrule
\multirow{4}{0.5cm}{R=16} & NMSE & 2.639 (1.131) & 2.185 (0.657) & \textbf{1.372} (0.384)\\& PSNR   &29.035 (2.384) & 29.708 (2.252)& \textbf{31.706} (2.118)\\
 & SSIM &0.901 (0.030) & 0.909 (0.020)& \textbf{0.937} (0.015)\\
 & HFEN & 0.274 (0.046)&0.248 (0.033) & \textbf{0.185} (0.028)\\
  \midrule
\multirow{4}{0.5cm}{R=24} & NMSE & 3.607 (1.226)&2.986 (0.811) &\textbf{2.148} (0.607) \\& PSNR &27.584 (2.157) & 28.320 (2.079) & \textbf{29.757} (2.110)\\
 & SSIM & 0.876 (0.035)& 0.886 (0.029) & \textbf{0.913} (0.022)\\
 & HFEN & 0.338 (0.060)& 0.305 (0.064) & \textbf{0.247} (0.047)\\
    \bottomrule
  \end{tabular}}
\end{stable}

\begin{sfigure}
\centering
\includegraphics[width=0.8\textwidth]{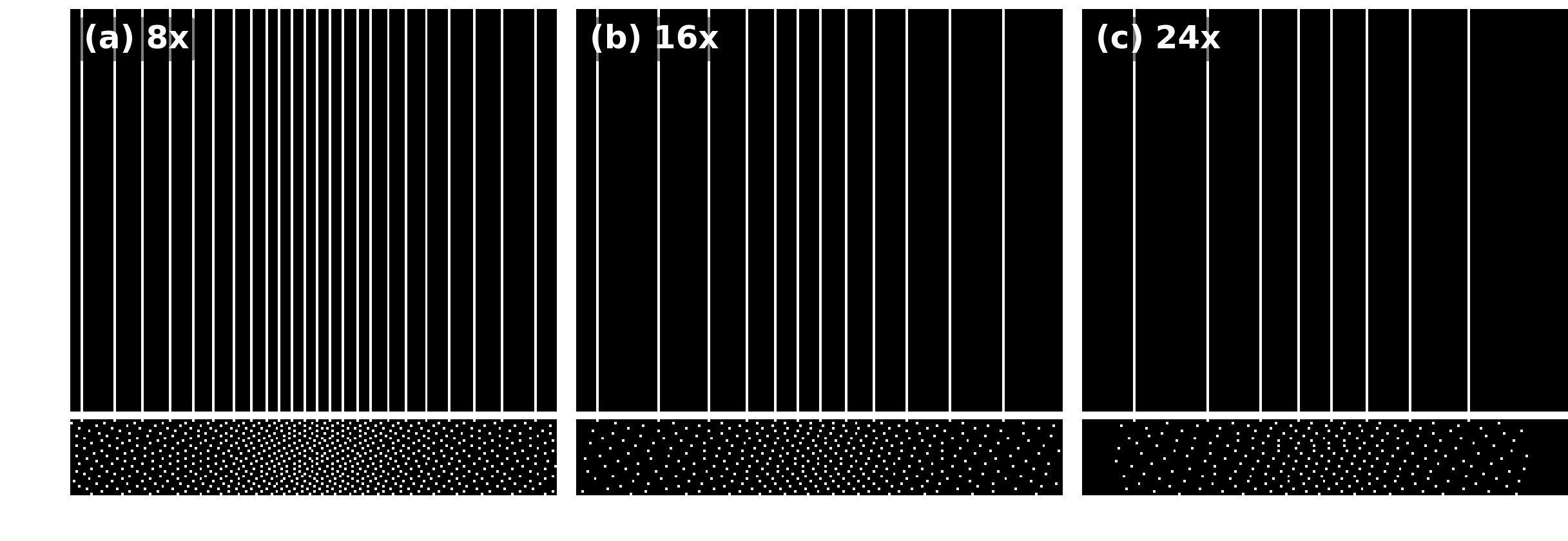}%
\caption{Examples of the VISTA undersampling patterns for acceleration rates 8, 16, and 24. Top figures show the undersampling patterns in $k$-space, and the bottom figures show the undersampling patters in $k$-$t$ space.}%
\label{sfig:masks}
\end{sfigure}

\begin{sfigure}
\centering
\includegraphics[width=.9\linewidth]{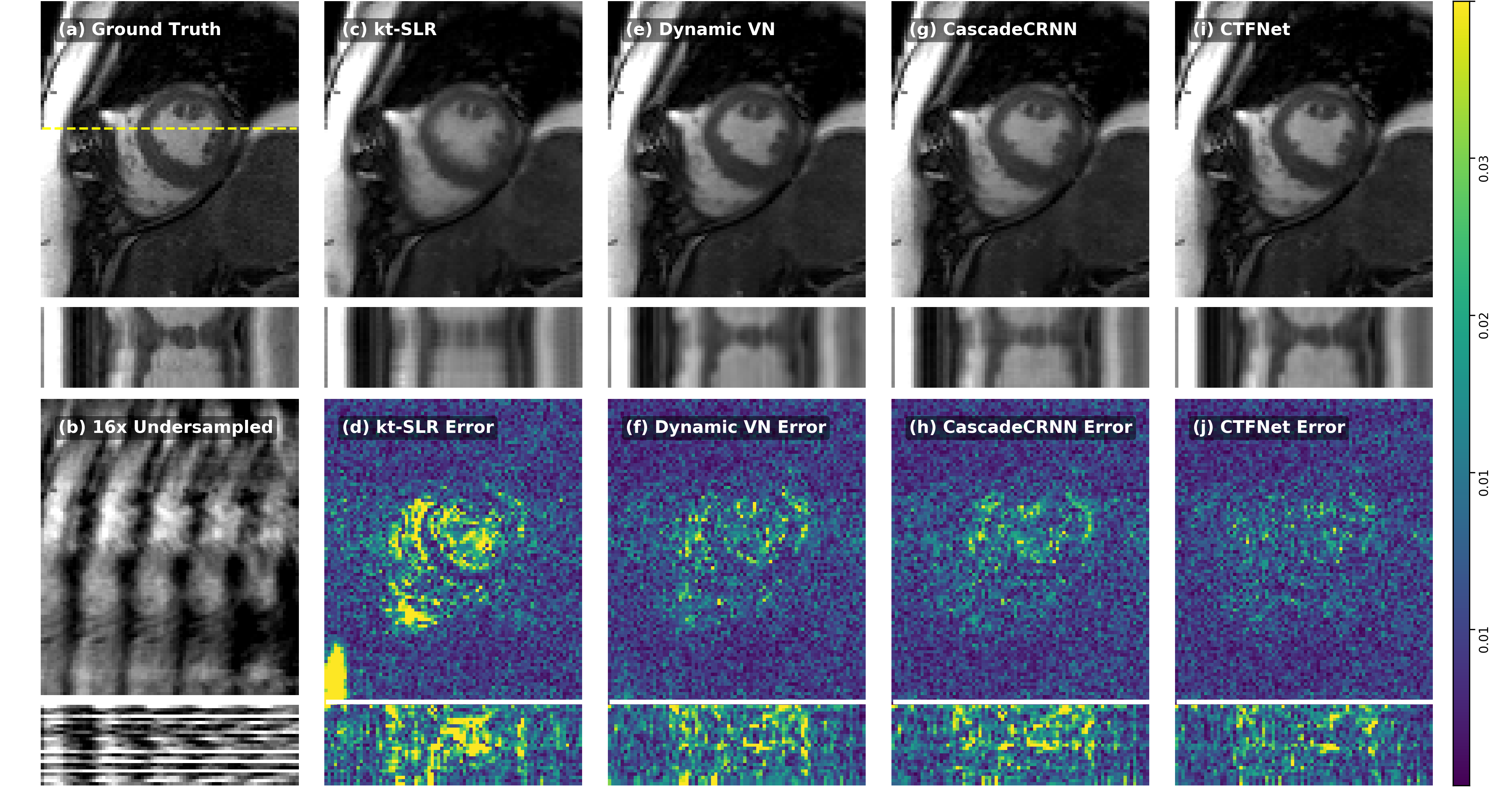}
% \vspace{2mm}
% \includegraphics[width=.9\linewidth]{figures/Pat17_ac16_im8_ES.png}
\vspace{2mm}
\includegraphics[width=.9\linewidth]{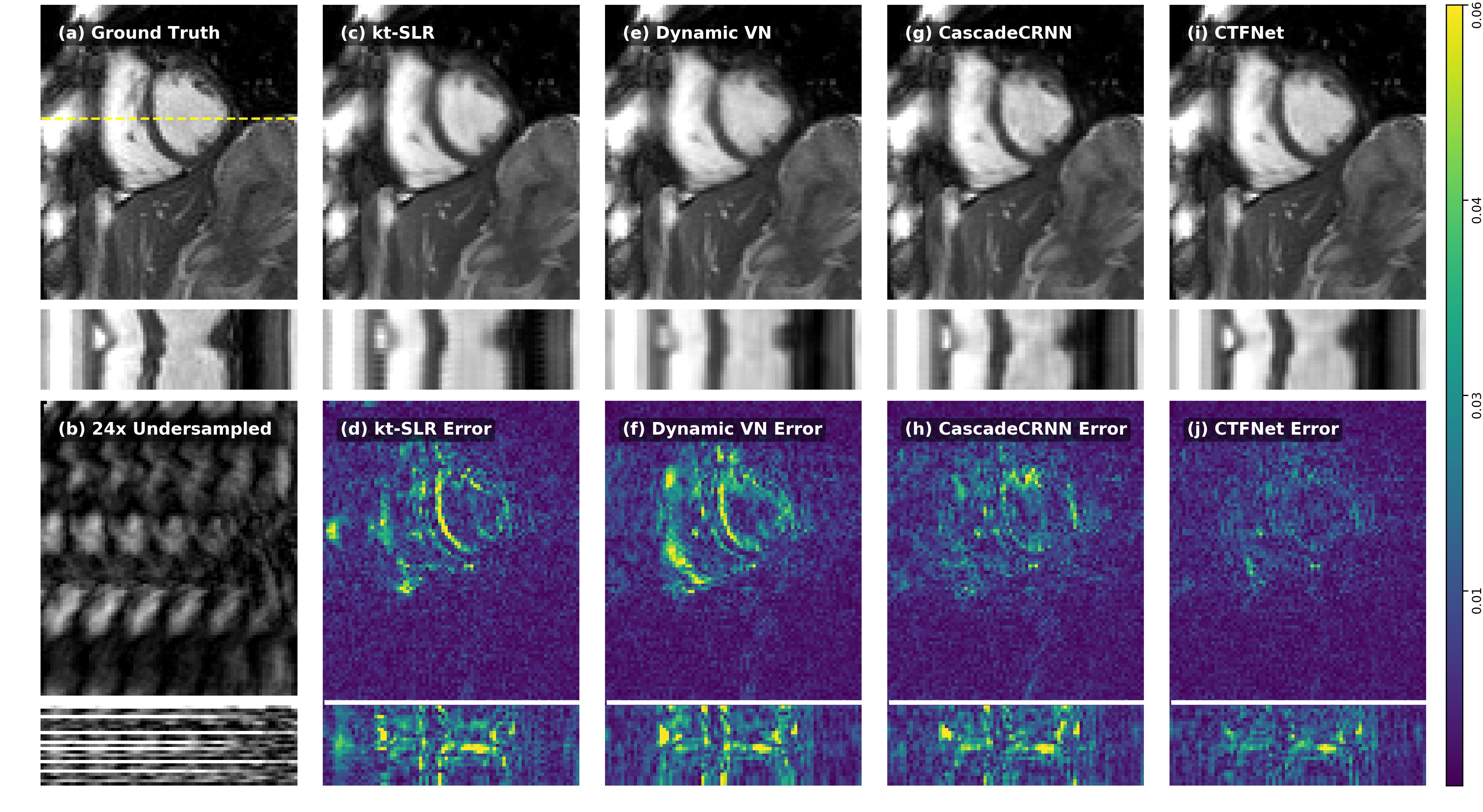}
% \vspace{2mm}
% \includegraphics[width=.9\linewidth]{figures/Al_ac24_im7_ES.png}
\caption{Qualitative comparison results of different methods on spatial and temporal dimensions with their error maps. Results are shown for undersampling rates R=16 of a patient (top) and R=24 of a healthy subject (bottom) on Dataset A at diastolic frames.}
\label{sfig:comparison_study}
\end{sfigure}

\begin{sfigure}
\centering
\includegraphics[width=0.6\linewidth]{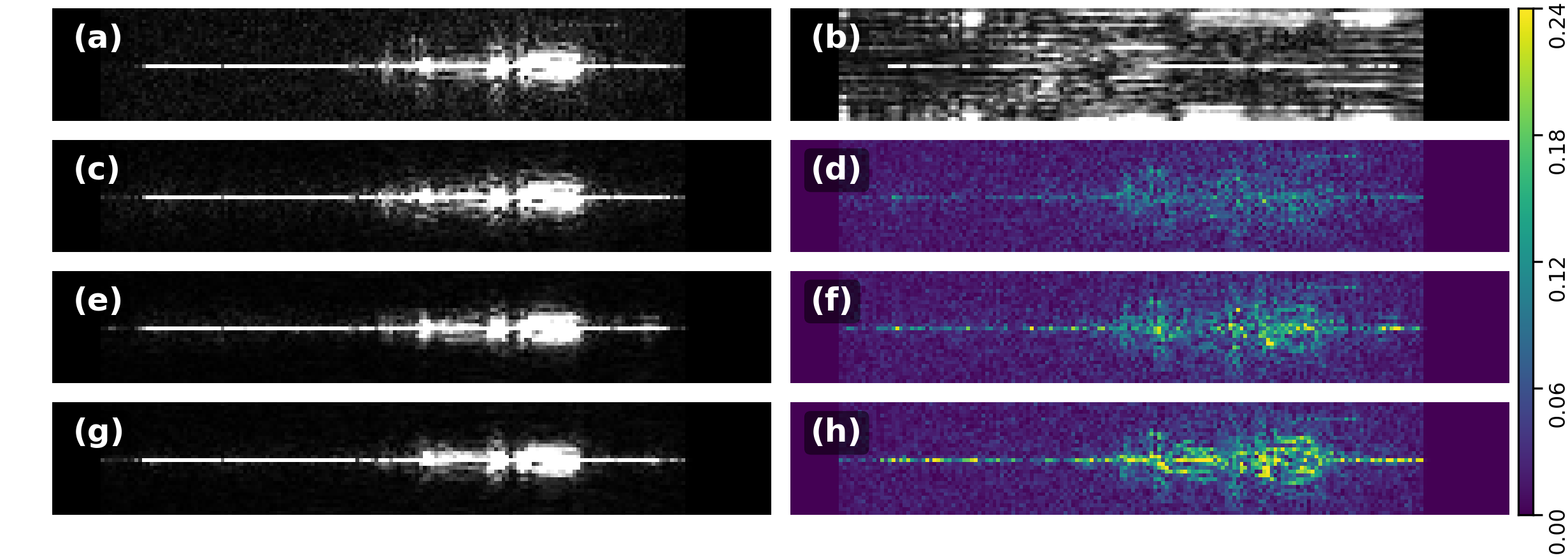}
\caption{$x$-$f$ reconstructions of CTFNet under different acceleration rates (R) with their error maps on dataset B. (a) Fully sampled signal (b) Undersampled example by R=16 (c) (d) $x$-$f$ reconstruction from R=8 (e) (f) $x$-$f$ reconstruction from AF R=16 (g) (h) $x$-$f$ reconstruction from AF R=24. }
\label{sfig:xf_comparison}
\end{sfigure}

\begin{sfigure}
     \centering
     \includegraphics[width=\linewidth]{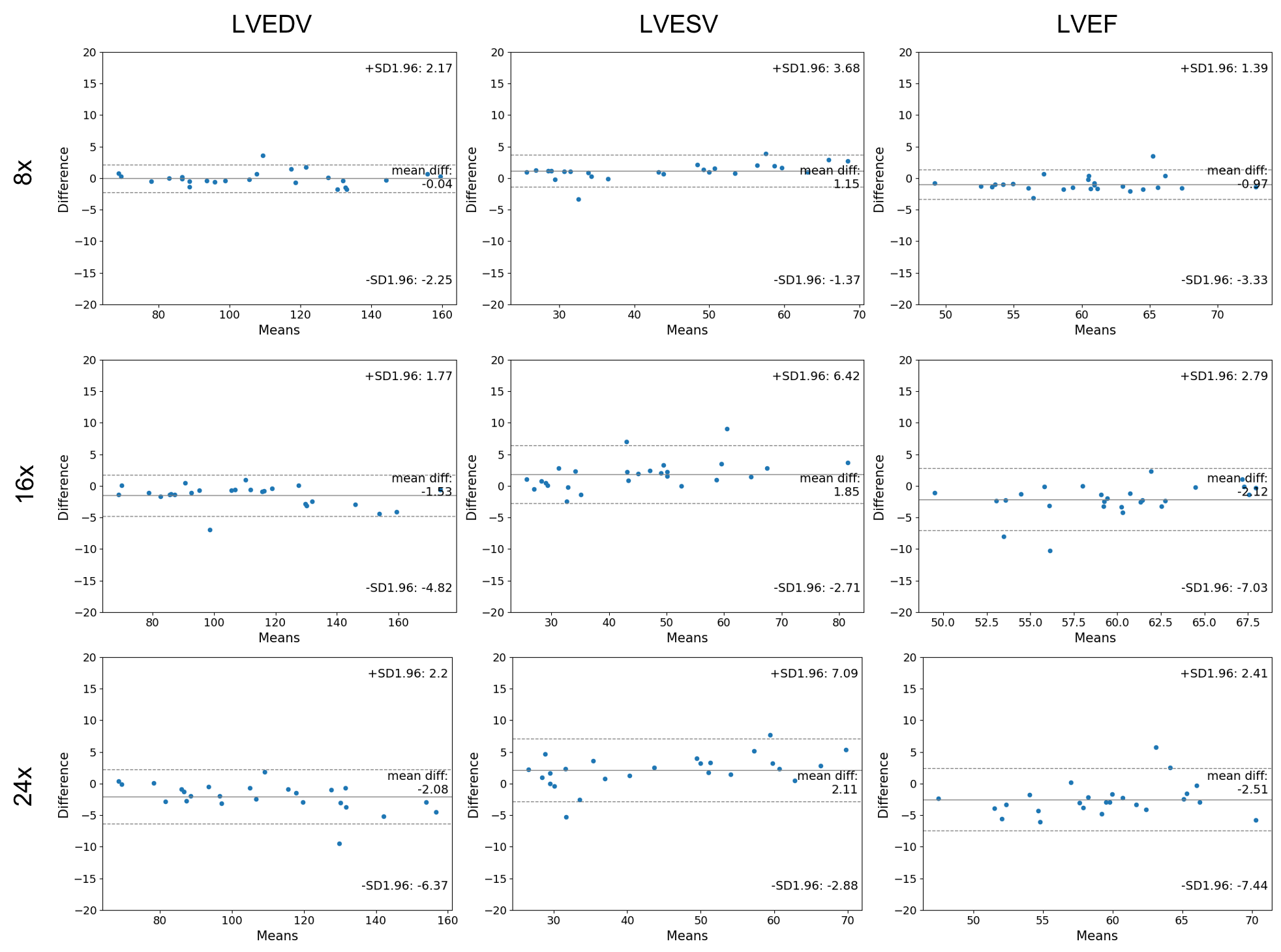}
     \caption{Bland-Altman plot for the LVEDV, LVESV and LVEF on the reference data and accelerated data (R=8, 16 and 24). The volumes were extracted by using an automated DL-based segmentation algorithm.}
     \label{sfig:baplot}
\end{sfigure}

\end{document}